\documentclass{article}

\usepackage{color}
\usepackage{geometry}
\usepackage{graphicx}
\usepackage[section]{placeins}
\usepackage{subfig}
\usepackage{hyperref}
\usepackage{stmaryrd}
\usepackage{amsmath,amssymb,amsfonts,amsthm,textcomp}
\usepackage{siunitx}
\usepackage[sort&compress,numbers,square]{natbib}
\usepackage{pgfplots}
\usepackage{authblk}
\pgfplotsset{every axis/.append style={line width=1pt}, every tick/.style={black}, compat=newest, grid style={dashed,gray}}

\newcommand{\bfj}{\boldsymbol{\jmath}}

\newcommand{\bfn}{\boldsymbol{n}}

\newcommand{\bfq}{\boldsymbol{q}}

\newcommand{\bft}{\boldsymbol{t}}
\newcommand{\bfu}{\boldsymbol{u}}
\newcommand{\bfv}{\boldsymbol{v}}

\newcommand{\bfx}{\textbf{x}}

\newcommand{\bfM}{\boldsymbol{M}}
\newcommand{\bfN}{\boldsymbol{N}}

\newcommand{\bfV}{\boldsymbol{V}}

\newcommand{\bfX}{\textbf{X}}

\newfont{\tenbfsl}{cmbxti9 scaled 1200}

\newcommand{\bfvarphi}{\boldsymbol{\varphi}}

\newcommand{\bfxi}{\boldsymbol{\xi}}

\newcommand{\bfchi}{\boldsymbol{\chi}}

\newcommand{\R}{_{\text{\tiny R}}}

\newcommand{\da}{\, \text{d}a}
\newcommand{\dx}{\, \text{d} \textbf{x}}
\newcommand{\dv}{\, \text{d}v}

\newcommand{\dvr}{\, \text{d}v_{\text{\tiny R}}}
\newcommand{\dar}{\, \text{d}a_{\text{\tiny R}}}
\newcommand{\dxr}{\, \text{d} \textbf{X}}

\newcommand{\be}{\begin{equation}}
\newcommand{\ee}{\end{equation}}

\newcommand{\calB}{\mathcal{B}}
\newcommand{\calC}{\mathcal{C}}

\newcommand{\calE}{\mathcal{E}}

\newcommand{\calP}{\mathcal{P}}

\newcommand{\calS}{\mathcal{S}}

\newcommand{\calW}{\mathcal{W}}

\newcommand{\bdy}{\calB}

\newcommand{\Def}{\vs\stackrel{\rm def}{=}\vs}

\newcommand{\trans}{\scriptscriptstyle\mskip-1mu\top\mskip-2mu}

\newcommand{\grad}{\hbox{\rm grad}\mskip2mu}

\newcommand{\sym}{\hbox{\rm sym}\mskip2mu}

\newcommand{\Divr}{\hbox{\rm Div}\mskip2mu}
\newcommand{\Divc}{\hbox{\rm div}\mskip2mu}

\newcommand{\prt}{\calP}

\newcommand{\vs}{\mskip2mu}

\newfont{\tenbbb}{msbm10}%
\newfont{\svnbbb}{msbm8}%

\newcommand{\xis}{\xi_{\scriptscriptstyle\calS}}

\newtheoremstyle{slplain}
  {.5\baselineskip\@plus.2\baselineskip\@minus.2\baselineskip}
  {.5\baselineskip\@plus.2\baselineskip\@minus.2\baselineskip}
  {\slshape}
  {}
  {\bfseries}
  {.}
  { }
  {}

\theoremstyle{slplain}

\begin{document}

\title{A continuum theory for mineral solid solutions undergoing chemo-mechanical processes}

\author[1]{Santiago P. \ Clavijo}
\author[2]{Luis Espath}
\author[3]{Adel Sarmiento}
\author[4,5,6]{Victor M. \ Calo}
\affil[1]{\footnotesize Ali I. Al-Naimi Petroleum Engineering Research Center, King Abdullah University of Science and Technology (KAUST), Thuwal, 23955-6900, Saudi Arabia.}
\affil[2]{Department of Mathematics, RWTH Aachen University, Pontdriesch 14-16, 52062 Aachen, Germany.}
\affil[3]{Mathematics, Mechanics, and Materials Unit, Okinawa Institute of Science and Technology (OIST), 1919-1 Tancha, Onna, Kunigami District, Okinawa 904-0495, Japan }
\affil[4]{School of Earth and Planetary Sciences, Curtin University, Kent Street, Bentley, Perth, WA 6102, Australia.}
\affil[5]{Curtin Institute for Computation, Curtin University, Kent Street, Bentley, Perth, WA 6102, Australia.}
\affil[6]{Mineral Resources, Commonwealth Scientific and Industrial Research Organisation (CSIRO), 10 Kensington, Perth, WA 6152, Australia.}

\date{}
\maketitle

\begin{abstract}
\noindent
Recent studies on metamorphic petrology as well as microstructural observations suggest the influence of mechanical effects upon chemically active metamorphic minerals. Thus, the understanding of such a coupling is crucial to describe the dynamics of geomaterials. In this effort, we derive a thermodynamically-consistent framework to characterize the evolution of chemically active minerals. We model the metamorphic mineral assemblages as a solid-species solution where the species mass transport and chemical reaction drive the stress generation process. The theoretical foundations of the framework rely on modern continuum mechanics, thermodynamics far from equilibrium, and the phase-field model. We treat the mineral solid solution as a continuum body, and following the Larch\'e and Cahn network model, we define displacement and strain fields. Consequently, we obtain a set of coupled chemo-mechanical equations. We use the aforementioned framework to study single minerals as solid solutions during metamorphism. Furthermore, we emphasise the use of the phase-field framework as a promising tool to model complex multi-physics processes in geoscience. Without loss of generality, we use common physical and chemical parameters found in the geoscience literature to portrait a comprehensive view of the underlying physics.  Thereby, we carry out 2D and 3D numerical simulations using material parameters for metamorphic minerals to showcase and verify the chemo-mechanical interactions of mineral solid solutions that undergo spinodal decomposition, chemical reactions, and deformation.
\end{abstract}

\section{Introduction}

When considering a deformable medium, chemical reactions may affect the solid's strength and its mechanical properties. Analogously, high mechanical strength may restrict either the volumetric shrinkage or swelling associated with the local volume changes caused by the chemical processes. Therefore, the chemical processes, associated with mass transport and chemical reactions, induce volume changes that lead to stresses around the reaction site. 

Finding innovative ways of approaching the modelling of solids is an essential open research topic in science and engineering. For instance, areas such as material science and geoscience are continually searching for new models that allow to improve the properties of materials or to understand the formation of mineral assemblages, which directly relate to solids undergoing chemical processes. In particular, metamorphic petrologists report the reciprocal chemo-mechanical responses of minerals during metamorphism \cite{milke2009matrix,tajvcmanova2015grain,hobbs2016does,moulas2013problem,vrijmoed2015thermodynamic,wheeler2014dramatic,zhong2017coupled}. A variety of study cases of this coupling ranges from grain-scale pressure variations in high-temperature metamorphic rocks to the proper definition of pressure in order to define P/T conditions of mineral assemblages during a metamorphic cycle. Without loss of generality, we use the aforementioned framework to study the tempo-spatial variations of stress-assisted volume changes.

The description of solidity and its properties is crucial to describe the physical and chemical responses of solids accurately. Gibbs' comprehensive study set the foundations for the thermodynamical properties of solids~\cite{ gibbs1878art}. However, Gibbs' solid model does not quantify the internal adjustment caused by compositional changes since the solid-state diffusion concept did not exist in his time. Herein, we seek to model multicomponent elastic solids that allow for changes in composition while remaining in the solid-state, and particularly, the impact of compositional changes on stress generation~\cite{ gurtin2010mechanics, dal2015computational, miehe2016phase, tsagrakis2017thermodynamic}. Larch\'e and Cahn introduced the equilibrium conditions for deformable bodies, which change composition as a result of chemical processes~\cite{ larche1973linear, larche1978nonlinear, larche1978thermochemical}. For instance, dissolution and precipitation at solid-fluid interfaces change the chemical composition of the solid, which in turn induce stresses associated with volume changes. Larche-Cahn's approach models the solid as a network, which allows us to define the stress-strain relations. A solid network can be, for example, the unit cell of the crystalline structure of a mineral, which arranges the atoms in a systematic and repeating pattern. Thus, the network model of Larch\'e and Cahn adequately describes a multicomponent solid.  

The outline of this work is as follows. Section~\S\ref{sec:fbt}, we present a coupled system of partial differential equations to model mass transport and deformation processes. Finally, section~\S\ref{sc:dimensionless} covers the dimensionless forms of the coupled system of chemo-mechanical equations for the multicomponent framework in conjunction with the dimensionless numbers of the coupling. Additionally, appendix~\S\ref{ap:framework} covers a thermodynamically-consistent treatment to the chemo-mechanical responses of the mineral solid solution. 

\section{Framework}\label{sec:fbt}
\subsection{Kinematics of Motion}

We introduce a system of partial differential equations to capture the evolution of a multicomponent elastic solid undergoing spinodal decomposition under reversible chemical reactions. In our framework, the deformations induced across the solid boundaries and compositional changes drive the stress generation process. Henceforth, we refer to this mechanism as stress-assisted volume changes.  We treat the solid as a continuum body that occupies an open subset \textbf{B} of the Euclidean space $\calE$. A time-dependent deformation field $\bfchi: \ $\textbf{B}$ \ \textsf{x} \ T  \rightarrow  \bdy_{t} \subset \calE$ describes the motion from a configuration \textbf{B} onto another configuration $\bdy_{t}$. We refer to $\textbf{B}$ as the reference configuration and to $\bfX$ \ as the particles in $\textbf{B}$. We set the reference configuration such that $\textbf{B}$ represents an undeformed state of the solid. The deformation field characterizes the kinematics of the motion of the particles in the body, and after deformation, it assigns to each material particle $\bfX$ \ at a given $t \in T$ a spatial particle $\bfx$ \ in the current configuration $\bdy_{t}$. Then, we express the deformation field as   
\begin{equation}
\bfx \Def \bfchi (\bfX, t) = \bfchi_{t} (\bfX),
\end{equation}
and abusing notation
\begin{equation}
\calB_{t} = \bfchi_{t} (\textbf{B}).
\end{equation}
The deformation field is invertible; namely, there exists an inverse deformation field $\bfchi^{-1}: \calB_{t} \ \textsf{x} \ T  \rightarrow  \ $\textbf{B}$ \ \subset \calE$ such that
\begin{equation}
\bfx = \bfchi_{t} (\bfchi^{-1} (\bfx, t)),
\end{equation}
which renders  
\begin{equation}
\bfX \Def \bfchi^{-1} (\bfx, t). 
\end{equation}
\subsection{Constitutive Equations}

Following \ref{ap:framework}, we derive a set of coupled chemo-mechanical equations. By using the Coleman-Noll procedure~\cite{ cimmelli2010generalized}, we find a set of constitutive equations as a pair for each kinematic process. We then rewrite~\eqref{eq:energyimb} following~\ref{eq:constfree.nchre} as
\be\label{eq:free.coleman.nchre}
\Bigg(\textbf{T}\R- \frac{\partial \hat{\psi}}{\partial \textbf{F}}\Bigg) \colon \dot{\textbf{F}} + \sum_{\alpha=1}^{n} \Bigg(\mu^{\alpha}\R - \pi^{\alpha} - \frac{\partial \hat{\psi}}{{\partial \varphi^{\alpha}\R}} \Bigg) \dot{\varphi^{\alpha}\R}+ \sum_{\alpha=1}^{n} \Bigg(\boldsymbol{\xi}^{\alpha}_{\text{\tiny R}} - \frac{\partial \hat{\psi}}{\partial {\nabla \varphi^{\alpha}\R}} \Bigg) \cdot \nabla \dot{\varphi^{\alpha}\R} - \sum_{\alpha=1}^{n} \left\{ \bfj^{\alpha}\R \cdot \nabla \mu^{\alpha}\R + \mu^{\alpha}\R s^{\alpha}_{\text{int}}\right\} \geq  0 .
\ee
Thereby, we find arbitrary values for $\dot{\textbf{F}}$, $\dot{\varphi}^{\alpha}\R$, $\nabla \dot{\varphi}^{\alpha}\R$, and $\nabla \mu^{\alpha}\R$ at a given time and position such that~\ref{eq:free.coleman.nchre} always hold. 

Following the notation and the definition for the Larch\'e--Cahn derivatives for both scalar and gradient fields as proposed by \cite{Cla21a}, the relative chemical potential $\mu^{\alpha}_{\text{\tiny R}\sigma}$ results from the Larc\'e-Cahn derivative as a consequence of incorporating the mass constraint given by~\eqref{eq:const.nchre}. According to Larch\'e-Cahn~\cite{ larche1978thermochemical}, the relative chemical potential expresses the chemical potential of $\alpha$-th species measured relative to the chemical potential of $\sigma$-th species. This definition entails that, for saturated systems, the mass constrain given by~\eqref{eq:const.nchre} must always hold. Analogously, the relative microforce $\boldsymbol{\xi}^{\alpha}_{\text{\tiny R}\sigma}$ emerges from the constraint imposed in the concentration gradients by~\eqref{eq:gradconst.nchre}. As a consequence, we rewrite~\eqref{eq:free.coleman.nchre} in the Larch\'e-Cahn sense the following terms: $\pi^{\alpha} := \pi^{\alpha}_{\sigma}$, $\mu^{\alpha} := \mu^{\alpha}_{\text{\tiny R} \sigma}$ and $\boldsymbol{\xi}^{\alpha}_{\text{\tiny R}} := \boldsymbol{\xi}^{\alpha}_{\text{\tiny R}\sigma}$ as well as the material mass fluxes $\bfj^{\alpha}\R := \bfj^{\alpha}_{ \text{\tiny R} \sigma}$ as all these quantities are expressed relative to the $\sigma$-th reference species. Thus, the energy imbalance is 
\be\label{eq:free.coleman2}
\Bigg(\textbf{T}\R- \frac{\partial \hat{\psi}}{\partial \textbf{F}}\Bigg) \colon \dot{\textbf{F}} + \sum_{\alpha=1}^{n} \Bigg(\mu^{\alpha}_{\text{\tiny R} \sigma} - \pi^{\alpha}_{\sigma} - \frac{\partial^{(\sigma)} \hat{\psi}}{{\partial \varphi^{\alpha}\R}} \Bigg) \dot{\varphi^{\alpha}\R} + \sum_{\alpha=1}^{n} \Bigg(\boldsymbol{\xi}^{\alpha}_{\text{\tiny R}\sigma} - \frac{\partial^{(\sigma)} \hat{\psi}}{\partial {\nabla \varphi^{\alpha}\R}} \Bigg) \cdot \nabla \dot{\varphi^{\alpha}\R} - \sum_{\alpha=1}^{n} \bfj^{\alpha}_{\text{\tiny R} \sigma} \cdot \nabla \mu^{\alpha}_{\R \sigma} \geq  0 .
\ee
The latter implies that the following relations must hold to keep consistency with the dissipation imbalance
\begin{subequations}
\begin{align}
&\textbf{T}\R = \frac{\partial \hat{\psi}}{\partial \textbf{F}}, \label{eq:consttraction}\\
&\pi^{\alpha}_{\sigma} = \mu^{\alpha}_{\text{\tiny R} \sigma}  - \frac{\partial^{(\sigma)} \hat{\psi}}{{\partial \varphi^{\alpha}\R}}, \label{eq:constmicroforce}\\
&\boldsymbol{\xi}^{\alpha}_{\text{\tiny R}\sigma} = \frac{\partial^{(\sigma)} \hat{\psi}}{\partial {\nabla \varphi^{\alpha}\R}}. \label{eq:constmicrostress}
\end{align} 
\end{subequations}

We use a logarithmic multi-well potential together with a multi-gradient-type potential for the chemical energy, that is, 
\be\label{eq:constfree12}
\hat{\psi}^{ch}(\bfvarphi\R, \nabla \bfvarphi\R) =  N_v k_B \vartheta \left( \sum _{\alpha=1} ^{n} \varphi^{\alpha}\R \ln \varphi^{\alpha}\R \right) + N_v \sum _{\alpha=1}^{n} \sum _{\beta=1} ^{n} \Omega ^{\alpha \beta} \varphi^{\alpha}\R \varphi^{\beta}\R + \dfrac{1}{2} \sum _{\alpha=1} ^{n} \sum _{\beta=1} ^{n} \Gamma ^{\alpha \beta} \, \nabla \varphi^{\alpha}\R \cdot \nabla \varphi^{\beta}\R.
\ee
This expression corresponds to the extension of the Cahn-Hilliard equation towards multicomponent systems~\cite{ cahn1958free, elliott1997diffusional}. The Ginzburg-Landau free energy  governs the dynamics of the phase separation process undergoing spinodal decomposition. In~\eqref{eq:constfree12}, $N_{v}$ is the number of molecules per unit volume, $k_B$ is the Boltzmann constant, and $\Omega^{\alpha \beta}$ represents the interaction energy between the mass fraction of the $\alpha$-th and $\beta$-th species. The interaction between the species $\alpha$ over $\beta$ is reciprocal, thus $\Omega^{\alpha \beta}$ is symmetric. The interaction energy is positive and is related to the critical temperature for each pair of species, $\vartheta_{c}^{\alpha \beta}$, (between the $\alpha$-th and $\beta$-th species). Following standard convention, we adopt that $\Omega^{\alpha \beta} = 0$ when $\alpha = \beta$ and $\Omega = 2k_{B}\vartheta_{c}^{\alpha \beta}$ when $\alpha \neq \beta$~\cite{ elliott1997diffusional,gurtin1989nonequilibrium,cahn1958free}. Furthermore, $\Gamma^{\alpha \beta} = \sigma^{\alpha \beta}\ell^{\alpha \beta}$ [force] (no summation implied by the repeated indexes) represents the magnitude of the interfacial energy between the $\alpha$-th and $\beta$-th species. The parameters $\sigma^{\alpha \beta}$ and $ l^{\alpha \beta}$ are the interfacial tension [force/length] and the interfacial thickness \footnote{This expression corresponds to the root mean square of the effective "interaction distance", as suggested by the work of Cahn and Hilliard~\cite{ cahn1958free}.} for each pair of species (between the $\alpha$-th and $\beta$-th species) [length], respectively. In~\cite{ cahn1958free}, the authors define the force $\Gamma^{\alpha \beta}$ as $N_{v}\Omega^{\alpha \beta} (\ell^{\alpha \beta})^{2}$.

Following~\cite{ miehe2016phase}, we assume the elastic solid behaves as a compressible neo-Hookean material whose elastic energy is given by
\begin{equation}\label{eq:elastic.energy.1}
\hat{\psi} ^{el} (\textbf{F}^{e}) = \dfrac{G}{2} \left[ \textbf{F}^{e} \colon \textbf{F}^{e} - 3 \right] + \dfrac{G}{\beta} \left[ (\text{det}  \textbf{F}^{e})^{-\beta}  - 1 \right],
\end{equation}
where $G$ and $\beta$ are material parameters that relate the shear modulus and the weak compressibility of the material. $\beta$ is a function of the Poisson ratio $\nu$ such that $\beta = 2 \nu / 1- 2\nu$. In line with treatments of thermoelasticity, we assume a multiplicative decomposition of the deformation gradient~\cite{ miehe2016phase}, that is, 
\begin{subequations}
\begin{align}\label{eq:elastic.energy.2}
\textbf{F}^{e} & = \textbf{F}^{\varphi} \textbf{F},\\
\textbf{F}^{\varphi}  & = \Bigg(1 + \sum _{\alpha=1} ^{n} \omega^{\alpha}(\varphi^{\alpha}\R - \varphi^{\alpha}_{\text{\tiny{R}} 0} ) \Bigg)^{-\frac{1}{3}} \textbf{I} ,\\
\textbf{F}^{\varphi}  & =  \text{J}^{-\frac{1}{3}}_\varphi\textbf{I}.
\end{align}
\end{subequations}
This expression suggests that as the local species concentrations change relative to their initial distribution, the solid must undergo elastic deformation. Moreover, the swelling material parameter $\omega_{\alpha}$ is related to the crystalline structure of the solid and its mechanical properties.

The evolution of the conserved field $\varphi^{\alpha}\R$ obeys a non-Fickian diffusion driven by the chemical potential differences between the species. We combine~\eqref{eq:constmicroforce} and~\eqref{eq:constmicrostress} using the balance of microforces~\eqref{eq:balmicro1} and the constitutive relation for the free energy~\eqref{eq:constfree.nchre} to express the relative chemical potential of the $\alpha$-th species as 
\be\label{eq:chemicalpot1.nchre}
\mu^{\alpha}_{\text{\tiny{R}} \sigma} = \frac{\partial^{(\sigma)} \hat{\psi}}{\partial \varphi^{\alpha}\R} - \Divr \frac{\partial^{(\sigma)} \hat{\psi}}{\partial \nabla \varphi^{\alpha}\R}-(\gamma^{\alpha}+\gamma^{\sigma}),
\ee
and therefore, 
\begin{equation}\label{eq:chemicalpot2.nchre}
\begin{split}
\mu^{\alpha}_{\text{\tiny{R}} \sigma} &= N_v k_{B} \vartheta \left( \ln \dfrac{\varphi^\alpha\R}{\varphi^\sigma\R} \right) + 2 N_v \sum_{\beta=1}^{n}  (\Omega^{\alpha \beta} - \Omega^{\sigma \beta}) \varphi^\beta\R\\ 
&- \sum _{\beta=1} ^N (\Gamma ^{\alpha \beta} - \Gamma ^{\sigma \beta}) \, \Divr \nabla \varphi^\beta\R - \frac{1}{3} \omega^{\alpha}_{\sigma} \text{J}_{\varphi}^{-1} \text{tr}[\textbf{T}\R \textbf{F}^{\trans}]-(\gamma^{\alpha}+\gamma^{\sigma}),
\end{split}
\end{equation}
where 
\begin{equation}
\omega^{\alpha}_{\sigma} = \omega^{\alpha} - \omega^{\sigma}.
\end{equation}
We define $p:= -\frac{1}{3}\text{tr}[\textbf{T}\R \textbf{F}^{\trans}]$ as the mechanical pressure and emphasize that this pressure modifies the mass transport rate. Therefore, for deformable bodies undergoing mass transport, this physical quantity alters the driving force of the chemical process. Furthermore, we define $p^{\alpha}_\varphi:= \omega^{\alpha}_{\sigma} \text{J}_{\varphi}^{-1} p$ as a mechanical pressure scaled by the local variation of $\alpha$-th species. Thus, we rewrite~\eqref{eq:chemicalpot2.nchre} as 
\begin{equation}\label{eq:chemicalpot2.nchre.p}
\begin{split}
\mu^{\alpha}_{\text{\tiny{R}} \sigma} &= N_v k_{B} \vartheta \left( \ln \dfrac{\varphi^\alpha\R}{\varphi^\sigma\R} \right) + 2 N_v \sum_{\beta=1}^{n}  (\Omega^{\alpha \beta} - \Omega^{\sigma \beta}) \varphi^\beta\R\\ 
&- \sum _{\beta=1} ^N (\Gamma ^{\alpha \beta} - \Gamma ^{\sigma \beta}) \, \Divr \nabla \varphi^\beta\R + p^{\alpha}_\varphi -(\gamma^{\alpha}+\gamma^{\sigma}),
\end{split}
\end{equation}

For convenience, we split the chemical potential $\mu^{\alpha}_{\text{\tiny{R}} \sigma}$ such that $\mu^{\alpha}_{\text{\tiny{R}} \sigma} = \mu^{\alpha}_{\varphi} + \mu^{\alpha}_{s} + p^{\alpha}_{\varphi}$. Thereby, 
\begin{equation}\label{eq:chemicalpot2.nchre.1}
\begin{aligned}
 \mu^{\alpha}_{\varphi}&= N_v k_{B} \vartheta \left( \ln \dfrac{\varphi^\alpha\R}{\varphi^\sigma\R} \right) + 2 N_v \sum_{\beta=1}^{n}  (\Omega^{\alpha \beta} - \Omega^{\sigma \beta}) \varphi^\beta\R-(\gamma^{\alpha}+\gamma^{\sigma}),\\
 \mu^{\alpha}_{s}&= -\sum _{\beta=1} ^N (\Gamma ^{\alpha \beta} - \Gamma ^{\sigma \beta}) \, \Divr \nabla \varphi^\beta\R,\\
 p^{\alpha}_{\varphi}&=-\frac{1}{3} \omega^{\alpha}_{\sigma} \text{J}_{\varphi}^{-1} \text{tr}[\textbf{T}\R \textbf{F}^{\trans}].
\end{aligned}
\end{equation}

The constitutive relation for the first Piola-Kirchhoff stress tensor is
\be\label{ref:piola}
\textbf{T}\R = G \text{J}_{\varphi}^{-1/3}[\textbf{F}^{e} - (\text{det} \textbf{F}^{e})^{-\beta} \textbf{F}^{e-\trans}].
\ee
We also consider the off-diagonal terms in the Onsager reciprocal relations and thus, we describe the species fluxes as
\begin{equation}\label{eq:mass.fluxes.nchre}
\bfj^{\alpha}_{\text{\tiny R} \sigma } \Def - \sum _{\beta=1} ^{n} \bfM^{\alpha \beta} \, J \textbf{C}^{-1} \nabla \mu^{\beta}_{\text{\tiny R} \sigma},
\end{equation}
where $\bfM^{\alpha \beta}$ are the Onsager mobility coefficients.  
We use the standard assumption that the mobility coefficients depend on the phase composition. In particular, we express this dependency in terms of the concentration of each species. We use the definition $\bfM^{\alpha \beta} = M^{\alpha\beta} \varphi^{\alpha}\R(\delta^{\alpha \beta} - \varphi^{\beta}\R)  \textbf{I}$ (no summation implied by the repeated indexes) where $\delta^{\alpha \beta}$ and $M^{\alpha\beta}$ are the Kronecker delta of dimension $n$ and the mobility coefficients~\cite{ elliott1997diffusional}, respectively.

\section{Dimensionless Form of the Chemo-Mechanical Equations}\label{sc:dimensionless}
Following \ref{ap:framework}, the coupled system of chemo-mechanical equations reads
\begin{equation}\label{eq:gov.eq.nchre.N}
\begin{aligned}
\dot{\varphi^{\alpha}\R} &= s^{\alpha}  - \Divr \bfj^{\alpha}_{\text{\tiny R} \sigma},\\
\bfj^{\alpha}_{\text{\tiny R} \sigma} &=  - \sum _{\beta=1} ^{n} \bfM^{\alpha \beta} \, J \textbf{C}^{-1} \nabla \mu^{\beta}_{\text{\tiny R} \sigma},\\
\mu^{\alpha}_{\text{\tiny R} \sigma} &=\frac{\partial^{(\sigma)} \psi}{\partial \varphi^\alpha\R} - \Divr \frac{\partial^{(\sigma)} \psi}{\partial (\nabla \varphi^\alpha\R)}-(\gamma^{\alpha}+\gamma^{\sigma}),\\
s^\alpha_{\text{int}} &= -\sum_{c=1}^{N_s}(\upsilon_{\alpha c}-\varpi_{\alpha c})(k_{c}^{+} \prod _{a=1} ^{n}  (\varphi^a\R) ^{\upsilon _{ac}}-k_{c}^{-} \prod _{a=1} ^{n}  (\varphi^a\R) ^{\varpi _{ac}}),\\
s^\alpha_{\mathrm{ext}} & = 0, \\
\textbf{0} &= \Divr \textbf{T}\R + \textbf{b},\\
\textbf{T}\R &= G \text{J}_{\varphi}^{-1/3}[\textbf{F}^{e} - (\text{det} \textbf{F}^{e})^{-\beta} \textbf{F}^{e-\trans}],
\end{aligned}
\end{equation}
with
\begin{equation}\label{eq:gov.eq.nchre.N}
\begin{aligned}
&\varphi^\alpha\R(\bfX,0)=\varphi^\alpha_0&\text{in}\,\textbf{P},\\
&\bfu(\bfX,0)=u_0&\text{in}\,\textbf{P},\\ \\
 &\text{subject to periodic boundary conditions.}
\end{aligned}
\end{equation}

Moreover, we introduce a free energy density $\psi_0 = 2 N_v k_{B} \vartheta$ together with a set of diffusion coefficients $\boldsymbol{D}^{\alpha \beta}$ such that 
\be
\boldsymbol{D}^{\alpha \beta} = \psi_0 M^{\alpha} \varphi^{\alpha}(\delta^{\alpha \beta} - \varphi^{\beta}).
\ee
To make dimensionless the energy densities, and the governing and constitutive equations, we define the following dimensionless variables
\be\label{eq:dimensiovar2123}
\overline{\textbf{u}} = u_0^{-1} \textbf{u}, \qquad \overline{\textbf{x}} = L_0^{-1}\textbf{x}, \qquad \overline{t} = D_0 l_0^2 L_0^{-4} t 
\ee
where $u_0$, $D_0$, and $l_0$ account for a reference deformation state, the diffusion coefficient, and the interface thickness of a reference species, respectively. We propose the following sets of scalar and vector dimensionless numbers for the multicomponent chemo-mechanical system, that is, 
\begin{equation}\label{eq:partwise.thermodynamical.laws.nchre}
\begin{aligned}
\omega^{\alpha}, & \ \ \overline{k}_{+}^{c} = k_{+}^c D_0^{-1} \ell^{-2}_0 L_0^4, \ \ \overline{k}_{-}^{c} = k_{-}^c D_0^{-1} \ell^{-2}_0 L_0^4, \ \ \overline{\vartheta}_c^{\alpha \beta} = \vartheta^{-1} \vartheta_{c}^{\alpha \beta},\\[4pt]
\overline{\ell}^{\alpha\beta} &= L_0^{-1}\ell^{\alpha \beta}, \ \ \overline{\psi} = \hat{\psi} \psi_0^{-1}, \ \ \overline{\sigma}^{\alpha \beta} = \sigma^{\alpha \beta}(\psi_0 L_0 )^{-1}, \ \ \beta, \ \ \overline{\textbf{b}} = G^{-1}\textbf{b},\\[4pt]
\overline{G} &= G \psi_0^{-1}, \ \ l = u_0 L_0^{-1}, \ \ \boldsymbol{\overline{D}}^{\alpha \beta} =  \boldsymbol{D}^{\alpha \beta} D_0^{-1} \ell_0^{-2} L_0^{2}, \ \ \overline{\gamma}^{\alpha} = \psi_0^{-1}\gamma^{\alpha}.
\end{aligned}
\end{equation}
By inserting the dimensionless quantities in~\eqref{eq:constfree12} and~\eqref{eq:elastic.energy.1}, we find the following dimensionless forms of the chemical energy
\be\label{eq:constfree123}
\begin{split}
\overline{\psi}^{ch}(\bfvarphi\R, \overline{\nabla} \bfvarphi\R) &=  \frac{1}{2} \left( \sum _{\alpha=1} ^{n} \varphi^{\alpha}\R \ln \varphi^{\alpha}\R \right) + \sum _{\alpha=1}^{n} \sum _{\beta=1} ^{n} \overline{\vartheta}_{c}^{\alpha \beta} \varphi^{\alpha}\R \varphi^{\beta}\R \\
&+ \frac{1}{2} \sum _{\alpha=1} ^{n} \sum _{\beta=1} ^{n} \overline{\sigma}^{\alpha \beta} \overline{\ell}^{\alpha \beta} \, \overline{\nabla} \varphi^{\alpha}\R \cdot \overline{\nabla} \varphi^{\beta}\R,
\end{split}
\ee
and the mechanical energy 
\begin{equation}\label{eq:elastic.energy.13}
\overline{\psi} ^{el} (\overline{\textbf{F}^{e}}) = \overline{G}\Bigg \{ \frac{1}{2}\left[ \overline{\textbf{F}^{e}} \colon \overline{\textbf{F}^{e}} - 3 \right] + \dfrac{1}{\beta} \left[ (\text{det}  \overline{\textbf{F}^{e}})^{-\beta}  - 1 \right]\Bigg \},
\end{equation}
where $\overline{\textbf{F}^{e}} = \text{J}_\varphi^{-1/3}(\textbf{I} + l \overline{\nabla} \overline{\bfu})$. 

Likewise, the dimensionless forms of the governing and constitutive equations read 
\be\label{eq:nchre.system}
\begin{aligned}
\frac{\partial \varphi^{\alpha}\R}{\partial \overline{t}} &= \overline{\nabla} \cdot \Bigg(\sum _{\beta=1} ^{n} \boldsymbol{\overline{D}}^{\alpha \beta}\overline{\bfM} \ \overline{\nabla} \overline{\mu}^{\beta}_{\text{\tiny{R}}\sigma}\Bigg) + \overline{s}^{\alpha},\\
\overline{\bfM} &= \text{det}(\textbf{I} + l\overline{\nabla} \overline{\bfu})(\textbf{I} + l \overline{\nabla} \overline{\bfu})^{-1} \textbf{I} (\textbf{I} + l\overline{\nabla} \overline{\bfu})^{-\trans},\\
\overline{\mu}^{\alpha}_{\text{\tiny{R}} \sigma} &= \frac{1}{2} \left( \ln \dfrac{\varphi^\alpha\R}{\varphi^\sigma\R} \right) + 2\sum_{\beta=1}^{n}  (\overline{\vartheta}_{c}^{\alpha \beta} - \overline{\vartheta}_{c}^{\alpha \beta}) \varphi^\beta\R - \sum _{\beta=1} ^N (\overline{\sigma} ^{\alpha \beta} \overline{\ell}^{\alpha \beta} - \overline{\sigma}^{\sigma \beta} \overline{\ell}^{\sigma \beta}) \, \overline{\Delta} \varphi^\beta\R \\ 
&-  \frac{1}{3} \omega^{\alpha \sigma} \text{J}_{\varphi}^{-1} \overline{G} \text{tr}[\overline{\textbf{T}}\R (\textbf{I} + l\overline{\nabla} \overline{\bfu})^{\trans}]-(\overline{\gamma}^{\alpha}+\overline{\gamma}^{\sigma}),\\
\overline{s}^\alpha_{\text{int}} &= -\sum_{c=1}^{n_s}\left\{(\upsilon^{c\alpha}-\varpi^{c\alpha})(\overline{k}^c_+ \prod _{a=1} ^{n} (\varphi^a\R) ^{\upsilon ^{ca}}-\overline{k}^c_- \prod _{a=1} ^{n} (\varphi^a\R) ^{\varpi ^{ca}})\right\},\\
\mathbf{0} & = \Divr \overline{\textbf{T}}\R + \overline{\textbf{b}},\\
\overline{\textbf{T}}\R &= \text{J}_{\varphi}^{-1/3}[\text{J}^{-\frac{1}{3}}_\varphi (\textbf{I} + l\overline{\nabla} \overline{\bfu}) - (\text{det} \text{J}^{-\frac{1}{3}}_\varphi (\textbf{I} + l\overline{\nabla} \overline{\bfu})^{-\beta} (\text{J}^{-\frac{1}{3}}_\varphi (\textbf{I} + l\overline{\nabla} \overline{\bfu}))^{-\trans}],
\end{aligned}
\end{equation}
subjected to the boundary conditions~\eqref{eq:gov.eq.nchre.N}.

\section{Numerical Simulations}\label{sc:numerics}

In this section, we present 2D and 3D simulations of the temporal evolution of single materials modelled as a solid solution composed of three phases $\mathcal{A}^1$, $\mathcal{A}^2$, and $\mathcal{A}^3$ to investigate their coupled chemo-mechanical interactions. In particular, the 2D simulation shows how interfacial interactions, together with a reversible chemical reaction between the phases, engender volumetric stresses as a result of local volume changes. The 3D simulation, on the other hand, studies a ripening mechanism. The interfacial interactions between the phases drive the phase separation process and allow for the Ostwald ripening and Gibbs-Thomson effects. We show the dimensionless temporal evolution of the dimensionless phases concentrations as well as the dimensionless displacements in each coordinate direction. By doing so, we seek to understand the interleaving between the physical and the chemical responses of the mineral solid solutions. 

\subsection{Reversible Chemical Reaction of Random Distributed Phases}\label{sc:reversible.stresses}
The reversible chemical reaction between the phases is 
\begin{equation}\label{eq:chemical.reaction_ex.nchre.s}
 \mathcal{A}^1 + \mathcal{A}^2 \mathop{\rightleftharpoons} _{k_-} ^{k_+} \, 2\mathcal{A}^3,
\end{equation}
where the stoichiometry vectors $\upsilon^{\alpha\beta}$ and $\varpi^{\alpha\beta}$ are given by
\begin{equation}\label{eq:chemical.sto.nchre.s}
\upsilon^{\alpha\beta}=(1,1,0),\qquad\text{and}\qquad\varpi^{\alpha\beta}=(0,0,2).
\end{equation}
We study the stress-assisted volume changes triggered by the chemical processes. Therefore, we do not drive the deformation by distorting the analysis domain. We set the external body forces and external microforces such that $\textbf{b} = \textbf{0}$ and $\gamma^{\alpha} = 0$, respectively. Moreover, we neglect all inertial effects. Consequently, the spatial velocity is $\bfv = \textbf{0}$. We assume that the three phases diffuse at the same rate. Therefore, we only consider one diffusion coefficient. In this simulation example, we set ${k_+} > {k_-}$. Thus, the forward chemical reaction occurs faster than the backward one. Our initial condition serves as the reference configuration, which we choose as an undeformed state of the body. The mass supply of each phase, captured by the reaction term $s^{\alpha}$, results solely from internal contributions as the system~\eqref{eq:chemical.reaction_ex.nchre.s} reacts. The initial spatial distribution of the phase concentrations is random such that $\varphi^{\alpha}$ takes values between $\varphi^{\alpha} \pm 0.05$ where we assume $\varphi^{\alpha}$ is $1/n$. We calculate the concentration of $\mathcal{A}^3$ following the mass constraint given by~\eqref{eq:const.nchre}. As mentioned before, we apply this mass constraint when we compute the relative quantities resulting from the Larch\'e--Cahn derivative. By doing so, we guarantee the consistency of the process. Furthermore, there is no mass flux at the solid boundaries. 

\begin{figure}
    \centering
    \includegraphics[scale=0.12]{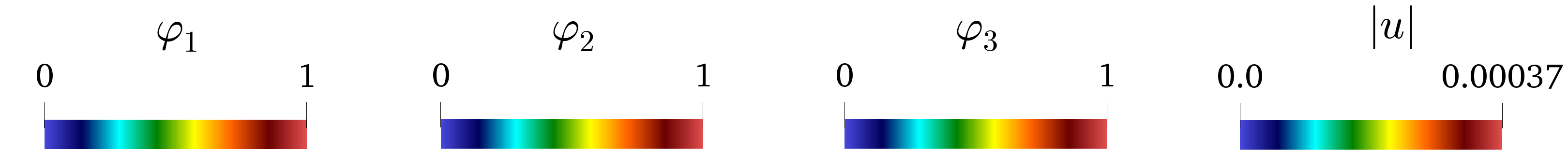}
    \subfloat[$t =0$]{{\includegraphics[scale=0.12]{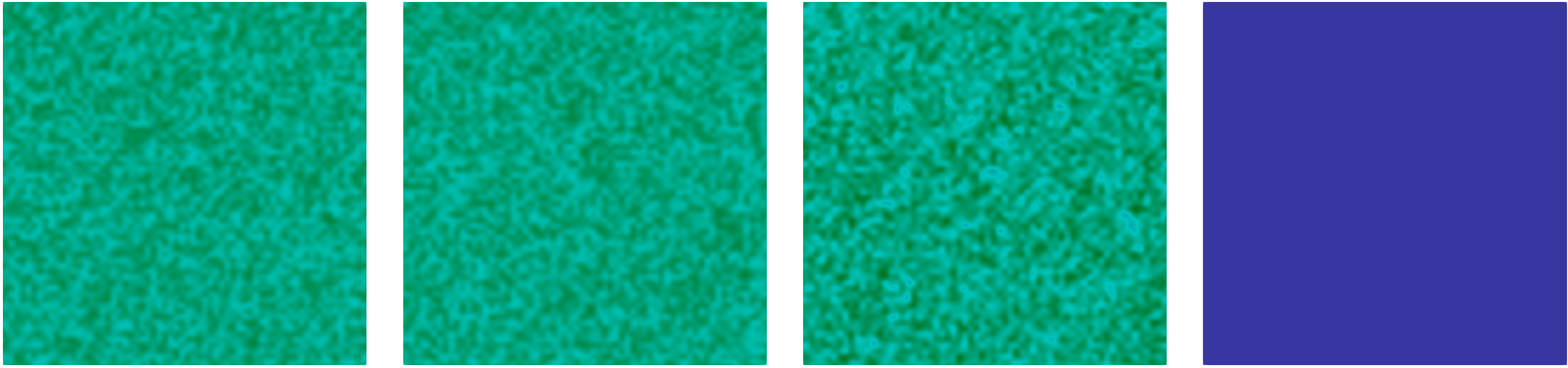} }}%
    \qquad
    \subfloat[$t = 4.04\times10^{-6}$]{{\includegraphics[scale=0.12]{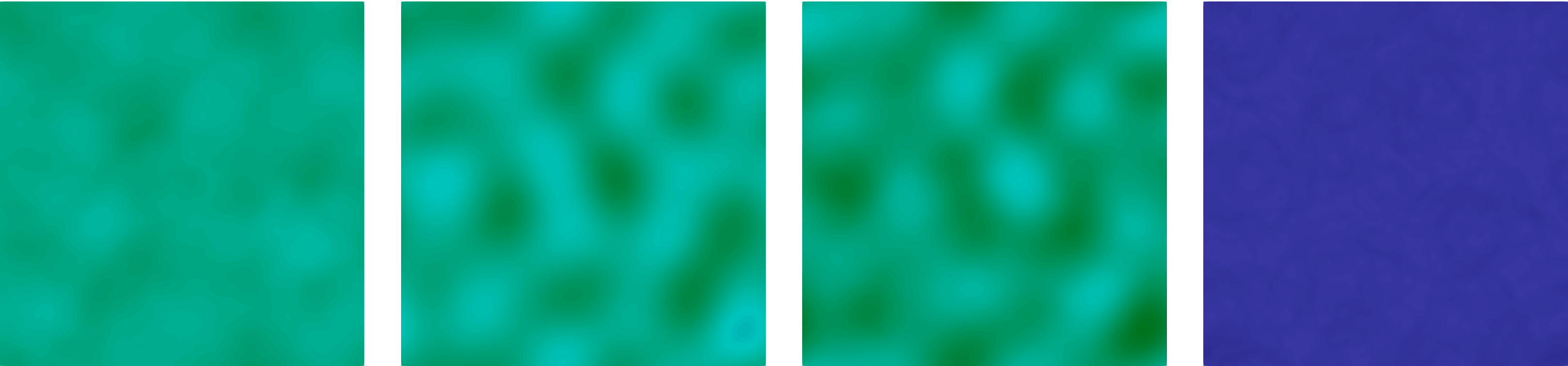} }}%
    \qquad
    \subfloat[$t =9.03\times10^{-6}$]{{\includegraphics[scale=0.12]{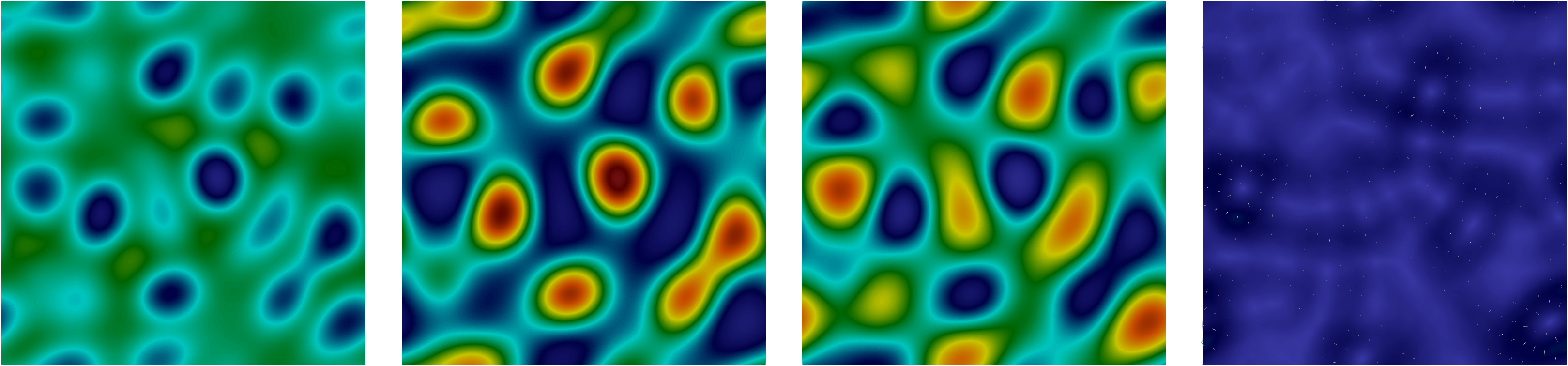} }}%
    \qquad
    \subfloat[$t=1.26\times10^{-5}$]{{\includegraphics[scale=0.12]{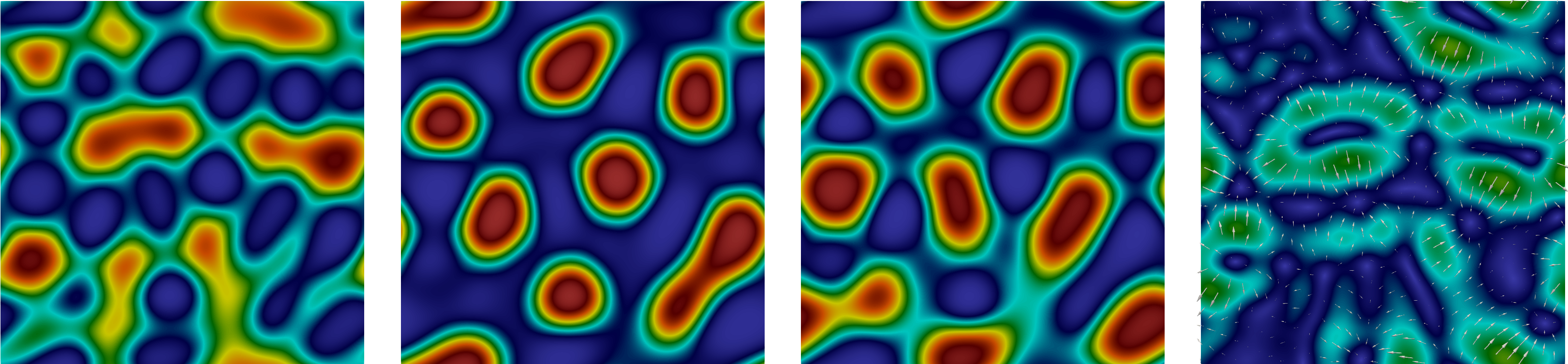} }}%
    \caption{Temporal evolution of the three phase fields together with the magnitud of the displacement vector and its direction at early stages.}
    \label{fig:evolution}%
\end{figure}

Figure~\ref{fig:evolution} (a) shows the spatial distribution of the initial phases concentrations in conjunction with the initial displacements. We set the parameters in the chemical energy such that we obtain a triple-well function. This function allows us to model the phase separation process. The reactive system seeks to minimize its global free energy. Thus, the reaction drives the given initial concentrations for the phases $\mathcal{A}^1$, $\mathcal{A}^2$, and $\mathcal{A}^3$ towards the concentrations at the well points.

Table~\ref{tb:pcparameters.nchre} summarises the parameters used to build up the dimensionless numbers as outlined in~\eqref{eq:partwise.thermodynamical.laws.nchre}.
\begin{table}[!htb]
\centering
\caption{Chemical and physical parameters that control the spinodal decomposition process}
\begin{tabular}{c c c}
\hline
Physical parameter & Value & Name \\
\hline
$\psi_0$ [\si{J.m^{-3}}] & $1 \times 10^{5}$ & Energy density\\
$L_0$ [\si{m}] & $10^{-6}$ & Domain length \\
$u_0$ [\si{m}] & $10^{-6}$ & Reference displacement\\
$\omega^{1}$ [-] & $0.0383$ & Swelling parameter phase 1\\
$\omega^{2}$ [-] & $0.0334$ & Swelling parameter phase 2\\
$\omega^{3}$ [-] & $0.0165$ & Swelling parameter phase 3\\
$G$ [GPa] & 40 & Shear modulus\\
$\beta$ [-] & 0.17 & Poisson's ratio \\
$\vartheta$ [\si{K}] & 727.0 & Absolute temperature\\
$\vartheta_{c}^{12}$ [\si{K}] & 800.0 & Critical temperature between phases 1 and 2  \\
$\vartheta_{c}^{13}$ [\si{K}] & 800.0 & Critical temperature between phases 1 and 3  \\
$\vartheta_{c}^{23}$ [\si{K}] & 800.0 & Critical temperature between phases 2 and 3  \\
$D$ [\si{m^2.s^{-1}}] &$10^{-20}$& Diffusion coefficient (same for all phases) \\
$k_+$ [\si{m^2.s^{-1}}] &$10^{-14}$ & Forward reaction rate \\
$k_-$ [\si{m^2.s^{-1}}] &$10^{-16}$ & Backward reaction rate \\
$\sigma$ [\si{J.m^{-2}}] & 0.817& Interfacial energy \\
$\ell$ [\si{m}] &$10^{-8}$& Interface thickness \\
$\gamma$ [-] &$0$& External microforce  (same for all phases)\\
\textbf{b} [\si{ms^{-2}}] &$\boldsymbol{0}$& Body force \\
\hline
\end{tabular}
\label{tb:pcparameters.nchre}
\end{table}
Hence,the diffusion matrix for each entry $\alpha$ and $\beta$ as well as the dimensionless numbers are given by 
\begin{equation}\label{eq:dimensionless.parameters.nchre}
\begin{gathered}
\boldsymbol{\bar{D}}^{\alpha\beta}=1\times10^{4}\varphi^{\alpha}(\delta^{\alpha\beta} - \varphi^{\beta})
\begin{bmatrix}
1 & 1 & 1 \\
0 & 1 & 1 \\
0 & 0 & 1
\end{bmatrix}\quad\forall\,1\le\alpha,\beta\le{n}, \quad
\bar{\vartheta}^{\alpha\beta}_c=
\begin{bmatrix}
0 & 1.100 & 1.100 \\
1.100 & 0 & 1.100 \\
1.100 & 1.100 & 0
\end{bmatrix}, \\
\bar{\sigma}^{\alpha\beta}\bar{\ell}^{\alpha\beta}=10^{-2}
\begin{bmatrix}
8.17 & 0 & 0 \\
0 & 8.17 &0 \\
0 & 0 & 8.17
\end{bmatrix},\quad
\upsilon^{\alpha\beta}=
\begin{bmatrix}
1 & 1 & 0
\end{bmatrix},\\
\varpi^{\alpha\beta}=
\begin{bmatrix}
0 & 0 & 2
\end{bmatrix},\quad
\bar{k}_+=0.01,\quad
\bar{k}_-=0.0001,\quad
\overline{G} = 4 \times 10^{5}
\end{gathered}
\end{equation}
where we choose $D_0 = D$ and $\ell_0 = \ell$ as the reference diffusion coefficient and interface thickness of a reference phase, respectively. 

The final system of coupled chemo-mechanical equations is
\begin{subequations}\label{eq:system.nchre.N}
 \begin{align}
&\dot{\varphi} ^1  = -k_+\varphi^1\varphi^2 + k_{-}(\varphi^3)^{2} - \Divr\bfj^1_{\text{\tiny R}3}, \\
&\dot{\varphi} ^2  = -k_+\varphi^1\varphi^2 + k_{-}(\varphi^3)^2- \Divr\bfj^2_{\text{\tiny R}3}, \\
&\Divr \textbf{T}\R = \textbf{0}.
\end{align}    
\end{subequations}
In~\eqref{eq:system.nchre.N}, we use the Larch\'e--Cahn derivative with $\mathcal{A}^3$ as the reference phase. We solve the system of partial differential equations in its primal form~\eqref{eq:weak.formulation.primal.material} and~\eqref{eq:weak.formulation.primal.material.l}. We state the problem as follows: find $\left\{ \varphi, \bfu\right\} \in \calC^2(\textbf{P})$ such that~\eqref{eq:nchre.system} given~\eqref{eq:gov.eq.nchre.N} subject to periodic boundary conditions up to the second derivative of $\varphi$, and $\bfu$ with respect to $\bfX$ in a square open region $\textbf{B} =(0,1)\times(0,1)$. We use PetIGA~\cite{ dalcin2016petiga}, a high-performance isogeometric analysis framework built on top of PETSc~\cite{ balay2019petsc}. We use a mesh with $64\times64$ elements of a polynomial degree $2$ and continuity $1$. 

We denote $\textit{H}^{2} $ as the Sobolev space of square integrable functions with square integrable first and second derivatives and $(\cdot,\cdot) _{\textbf{P}}$ as the $\textit{L}^{2}$ inner product over an arbitrary material part $\textbf{P}$ with boundary $\text{S}$. We multiply the Lagrangian version of the phases mass balance~\eqref{eq:massbalR2.nchre} by a test function $\varrho^{\alpha}$, which belongs to $\textit{H}^{2} $, using the definition for the material mass fluxes~\eqref{eq:mass.fluxes.nchre} and integrating by parts, the primal variational formulation can then be given by:
\begin{equation}\label{eq:weak.formulation.primal.material}
\begin{split}
(\varrho ^\alpha, \dot{\varphi ^\alpha}) _{\textbf{P}} = &(\varrho ^\alpha, s ^\alpha) _{\textbf{P}} - (\varrho ^\alpha, \bfj^{\alpha}_{\text{\tiny R}\sigma I,I}) _{\textbf{P}} \\
 = & (\varrho ^\alpha, s ^\alpha) _{\textbf{P}}  + (\varrho ^\alpha_{,I}, \bfj^{\alpha}_{\text{\tiny R}\sigma I}) _{\textbf{P}} - (\varrho ^\alpha, \bfj^{\alpha}_{\text{\tiny R}\sigma I} \bfN _I) _{\text{S}} \\ 
 = & (\varrho ^\alpha, s ^\alpha) _{\textbf{P}}  + (\varrho ^\alpha_{,I}, -\bfM^{\alpha\beta}  (\mu ^\beta _\varphi + \mu ^\beta _s + p^\beta_\varphi)_{,J} \textbf{C}^{-1}_{JI} J ) _{\textbf{P}}\\
 &- (\varrho ^\alpha, - \bfM^{\alpha\beta} (\mu ^\beta_{,J}) \textbf{C}^{-1}_{JI} J  \bfN _I) _{\text{S}} \\
 = & (\varrho ^\alpha, s ^\alpha) _{\textbf{P}} - (\varrho ^\alpha_{,I}, \bfM^{\alpha\beta} \mu ^\beta _{\varphi,J} \textbf{C}^{-1}_{JI} J) _{\textbf{P}} - (\varrho ^\alpha_{,I}, (\bfM^{\alpha\beta} \mu ^\beta _s)_{,J} \textbf{C}^{-1}_{JI} J) _{\textbf{P}} \\ %
 &+ (\varrho ^\alpha_{,I}, (\bfM^{\alpha\beta}_{,J}) \mu ^\beta _s \textbf{C}^{-1}_{JI} J) _{\textbf{P}} - (\varrho ^\alpha_{,I}, \bfM^{\alpha\beta} p^\beta _{\varphi,J} \textbf{C}^{-1}_{JI} J) _{\textbf{P}} \\
 &+ (\varrho ^\alpha, \bfM^{\alpha\beta} (\mu ^\beta_{,J})  \textbf{C}^{-1}_{JI} J \bfN _I) _{\text{S}} \\
= & (\varrho ^\alpha, s ^\alpha) _{\textbf{P}} - (\varrho ^\alpha_{,I }, \bfM^{\alpha\beta} \mu ^\beta _{\varphi,J} \textbf{C}^{-1}_{JI} J) _{\textbf{P}} + (\varrho ^\alpha_{,IJ} \textbf{C}^{-1}_{JI} J, \bfM^{\alpha\beta} \mu ^\beta _s) _{\textbf{P}} \\
& +(\varrho ^\alpha_{,I} (\textbf{C}^{-1}_{JI} J)_{,J}, \bfM^{\alpha\beta} \mu ^\beta _s)_{\textbf{P}}  + ( \varrho ^\alpha_{,I}, (\bfM^{\alpha\beta}_{,J}) \mu ^\beta _s  \textbf{C}^{-1}_{JI} J) _{\textbf{P}}\\
& - (\varrho ^\alpha_{,I}, \bfM^{\alpha\beta} p^\beta _{\varphi,J} \textbf{C}^{-1}_{JI} J) _{\textbf{P}} + (\varrho ^\alpha, \bfM^{\alpha\beta} (\mu ^\beta_{,J}) \bfN _I \textbf{C}^{-1}_{JI} J) _{\text{S}}.\\ 
&- (\varrho ^\alpha_{,I} \bfN_J \textbf{C}^{-1}_{JI} J, \bfM^{\alpha\beta} \mu^\beta _s) _{\text{S}}\\
\end{split}
\end{equation}

Furthermore, the weak formulation of the Lagrangian version of the linear momenta balance reads
\begin{equation}\label{eq:weak.formulation.primal.material.l}
(w_{i},\textbf{T}_{\text{\tiny R}iI} \bfN_{I})_{\text{S}} - (w_{i,I},\textbf{T}_{\text{\tiny R}iI})_{\textbf{P}} = 0,
\end{equation}
where we multiply~\eqref{eq:lm} by a test function $w_{i}$.

\begin{figure}[]
\footnotesize
   \centering
    \subfloat{{\includegraphics[scale=0.45]{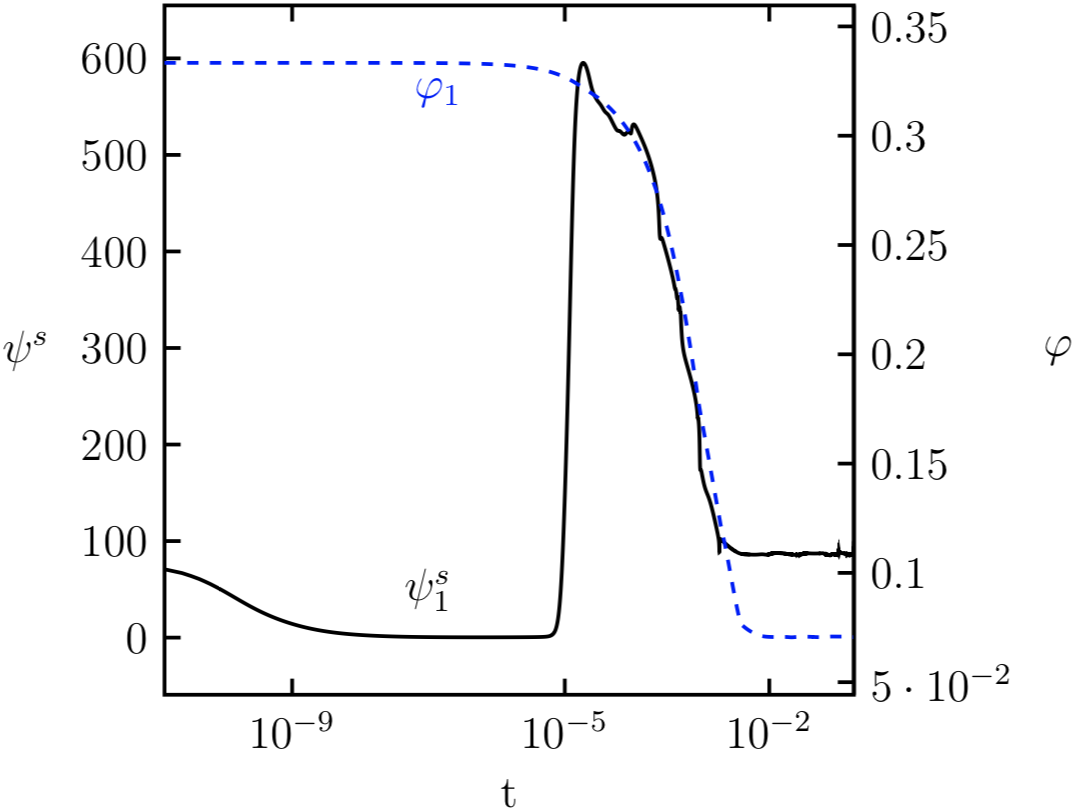}}}%
    \qquad
    \subfloat{{\includegraphics[scale=0.45]{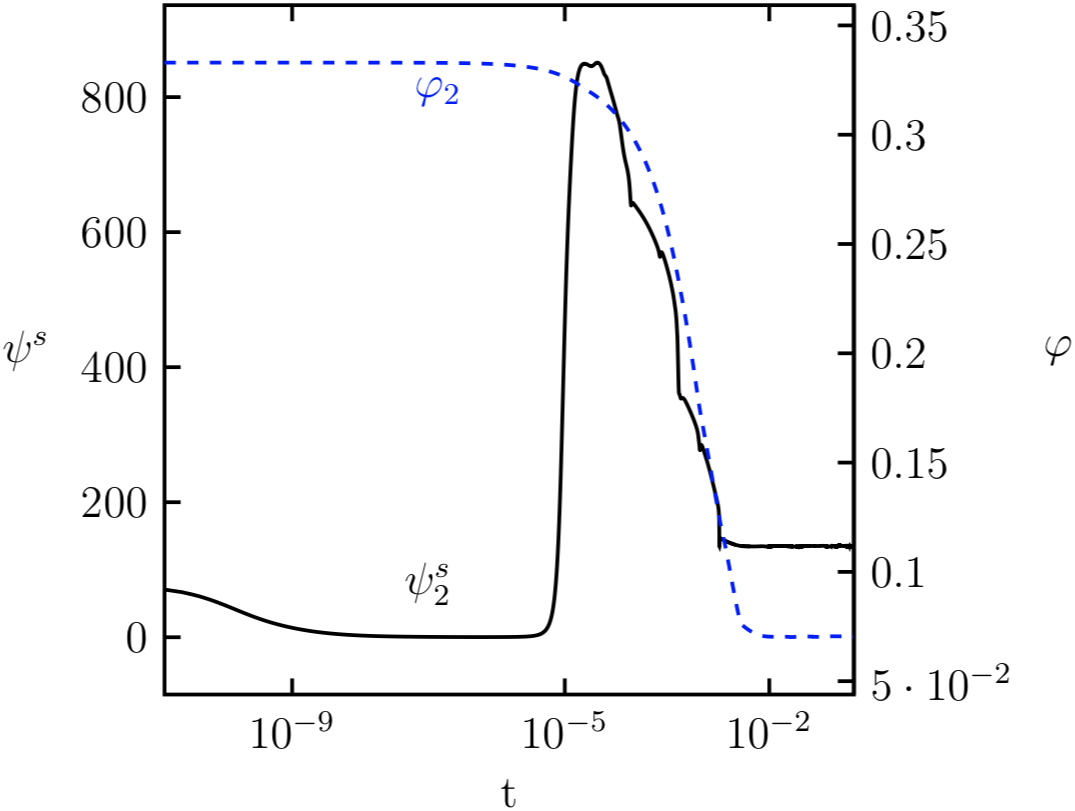}}}
    \qquad
   \subfloat{{\includegraphics[scale=0.45]{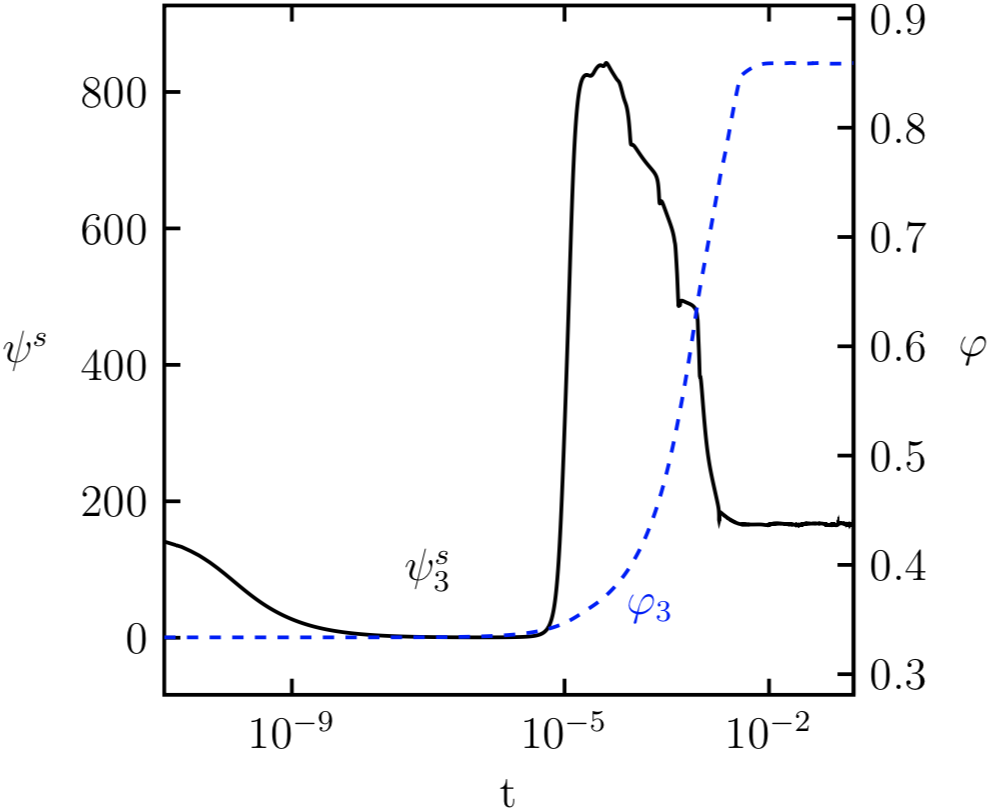}}}
    \caption{Interfacial energies for phases $\mathcal{A}^1$, $\mathcal{A}^2$, and $\mathcal{A}^3$ along with their masses}
    \label{fig:masses.nchre}
\end{figure}

At early stages Figure~\ref{fig:evolution} (b), $t < 4.04\times10^{-6}$, the solution goes through an initial spinodal decomposition. This spontaneous phase separation process occurs due to $\vartheta^{\alpha\beta}_c > \vartheta$. Otherwise, the mixture would only diffuse without unmixing. As the phases $\mathcal{A}^{i}$, $i=1,2,3$, diffuse as a result of their separation, the solid undergoes elastic deformation due to the mass transport. Analogously, the pressure $p^{\alpha}_{\varphi}$ alters the rate at which the phases diffuse. The deformation arises solely from mass transport since the reversible chemical reaction has no significant impact. From Figure~\ref{fig:masses.nchre}, we verify that the phases masses do not change substantially in the range $0 < t < 4.04\times10^{-6}$. As a consequence, there is no nucleation and growth of phases. With regards to the interfacial energies, in the range $0< t < 4.04\times10^{-6}$ the energies decrease gently up to a point, $t \simeq 1.0\times10^{-7}$, where the interfacial energies become constant. A small change in the interfacial energies means that either the phase separation has not evolved significantly or there is no substantial coarsening. Furthermore, at the early stages, the deformation is small as the displacements (see Figure~\ref{fig:evolution}) are not large since the phase separation has not evolved such that the phases are unmixed. As expected, the displacements $u_x$ and $u_y$ in the solid move following the mass transport. 

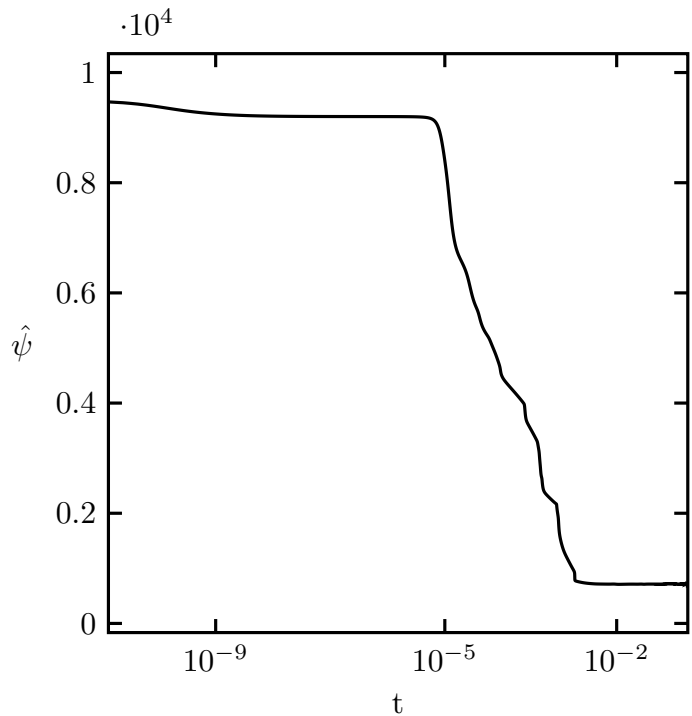
\begin{figure}[]
\centering
\begin{tikzpicture}[scale=1.2]
\begin{semilogxaxis}[xlabel=t,
				ylabel=$\hat{\psi}$,ylabel style={rotate=-90},
				xmax=1.742849e-01,xmin=1.342177e-11, 
				xtick={1e-9,1e-5,1e-2,1},
				width=8cm,height=8cm]
\addplot [color=black] table [x=time,y=energy] {results_nCHRe.dat};
\end{semilogxaxis}
\end{tikzpicture}
\caption{During the whole evolution, the free energy is monotonically-decreasing.}%
\label{fig:freeener.nchre}
\end{figure}

Later on, in the range between $4.04\times10^{-6} < t < 9.03\times10^{-6}$, the phase separation becomes prominent  as it allows to form spatial domains rich in each component (see Figure~\ref{fig:evolution} (c)). In particular, the phase $\mathcal{A}^1$ remains partially unmixed as there are no rounded inclusions with large concentration (see Figure~\ref{fig:evolution} (c)). On the contrary, for phases  $\mathcal{A}^2$ and $\mathcal{A}^3$, rounded inclusions with large concentrations appear. The deformation results from the mass transport itself as there is no substantial influence of the reversible chemical reaction (see Figure~\ref{fig:masses.nchre}). Larger displacements are collocated with the larger inclusions for $\mathcal{A}^2$ and $\mathcal{A}^3$. This collocation is a consequence of the mass flux towards these points as the inclusions grow. The enlargement associated with the growth of the inclusions induces deformation. $\textbf{F}^{\varphi}$ captures this behavior. The interfacial energies remain roughly constant, which implies that there is a balance between both the creating and disassemble of phases interfaces. As expected, in the time interval between $0< t < 9.03\times10^{-6}$, the tendency of the global free energy is monotonically decreasing as the system goes to a steady state of maximum entropy (see Figure~\ref{fig:freeener.nchre}). This free energy encompasses the contribution from both the chemical and the mechanical energies. At the early stages, as the evolution towards a steady-state goes on, the system favors phase separation. 

For instance, Figure~\ref{fig:evolution} (d) shows the evolution of small $\mathcal{A}^1$ inclusions in the range between $9.03\times10^{-6}< t < 1.26\times10^{-5}$. As suggested before, there is mass flux towards these points that allows the inclusion to grow. Consequently, there must be deformation associated with mass transport. Figure~\ref{fig:evolution} (d) also shows the larger displacements in the regions where the inclusions are. On the other hand, $\mathcal{A}^3$ inclusions are large and close enough to start merging. This phenomenon is associated with the minimization of the global free energy as the system reduces its interfacial energies. Nevertheless, when considering the system undergoing chemical reactions, the interfacial energies evolve according to the chemical reaction, which in this modeling example corresponds to a reversible chemical reaction. The growth of more $\mathcal{A}^3$ as a result of the reversible chemical reaction creates more $\mathcal{A}^3$ interface. Figure~\ref{fig:masses.nchre} shows an increase in $\mathcal{A}^3$ interfacial energy as well as its mass. On the other hand, the decomposition of $\mathcal{A}^3$ into $\mathcal{A}^1$ and $\mathcal{A}^2$ following~\eqref{eq:chemical.reaction_ex.nchre.s} must increase $\mathcal{A}^1$ and $\mathcal{A}^2$ masses, and their interfacial energies. However, this behavior does not last since the rate of creation of $\mathcal{A}^3$ is faster than the decomposition into $\mathcal{A}^1$ and $\mathcal{A}^2$. This is due to $k_{+} > > k_{-}$. 

\begin{figure}
 \centering  
   \includegraphics[scale=0.12]{figure2}
    \qquad 
    \subfloat[$t =5.64\times10^{-5}$]{{\includegraphics[scale=0.12]{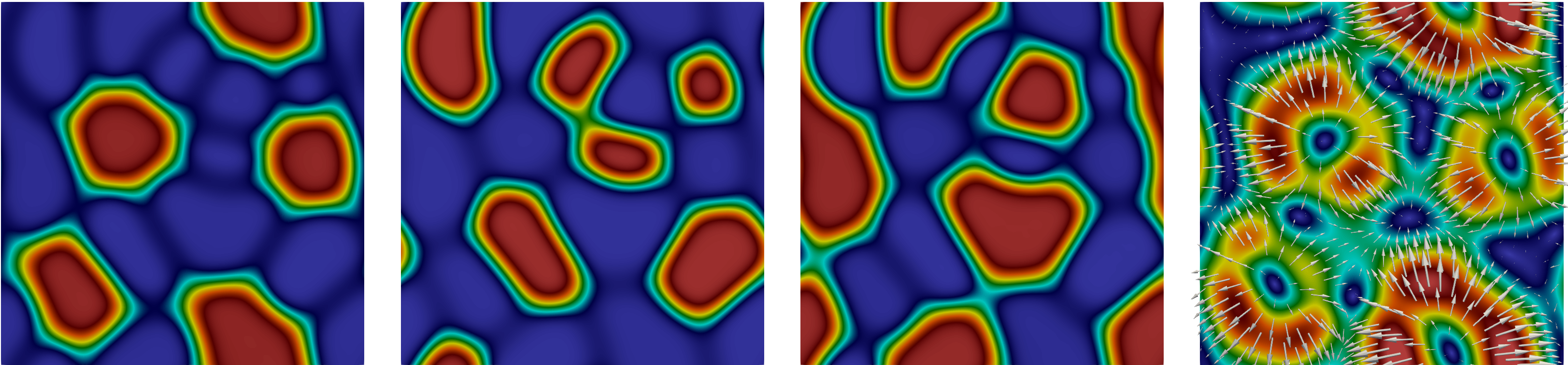} }}%
    \qquad
    \subfloat[$t=4.26\times10^{-4}$]{{\includegraphics[scale=0.12]{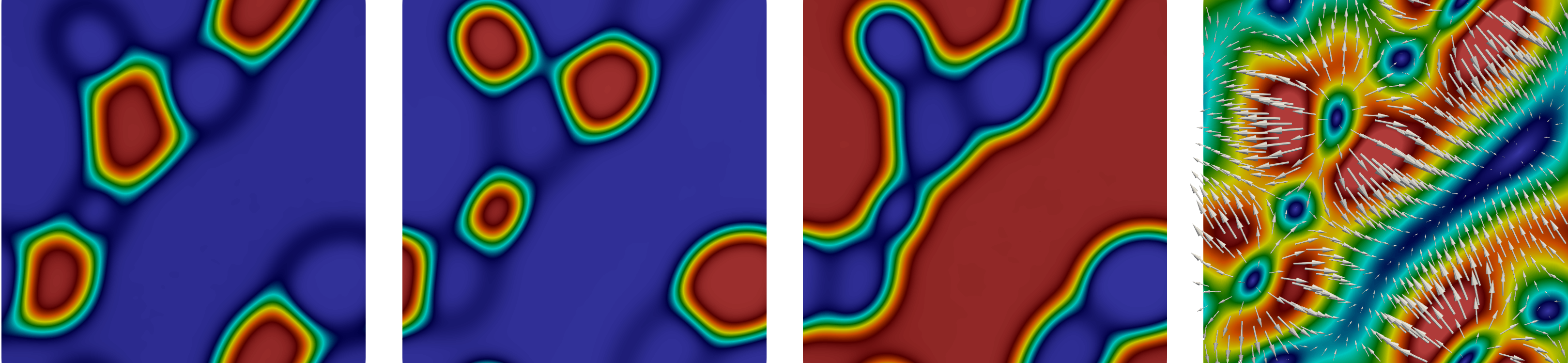} }}%
    \qquad
    \subfloat[$t=5.72\times10^{-4}$]{{\includegraphics[scale=0.12]{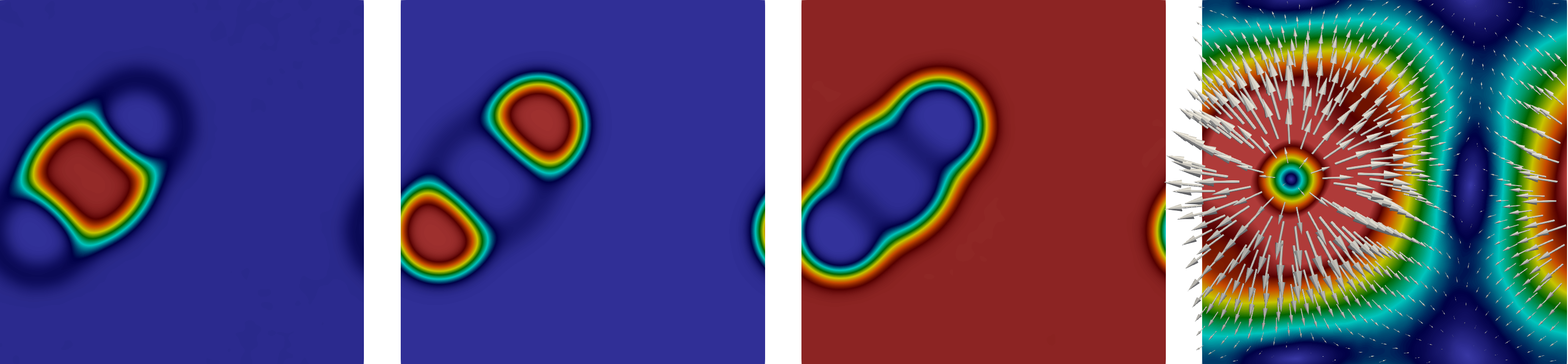} }}%
    \qquad
    \subfloat[$t=1.84\times10^{-3}$]{{\includegraphics[scale=0.12]{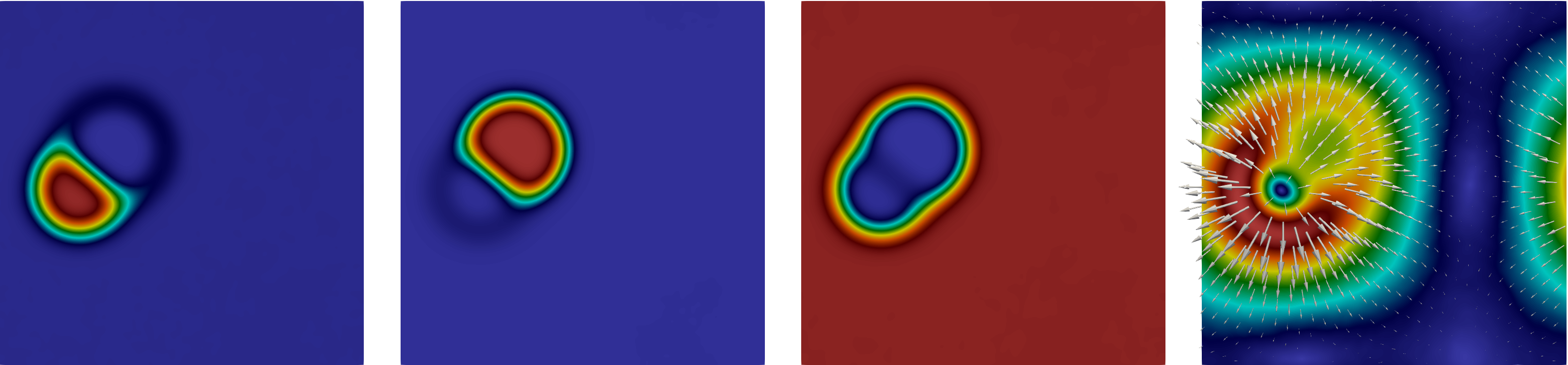} }}%
    \qquad
    \subfloat[$t > 10^{-2}$]{{\includegraphics[scale=0.12]{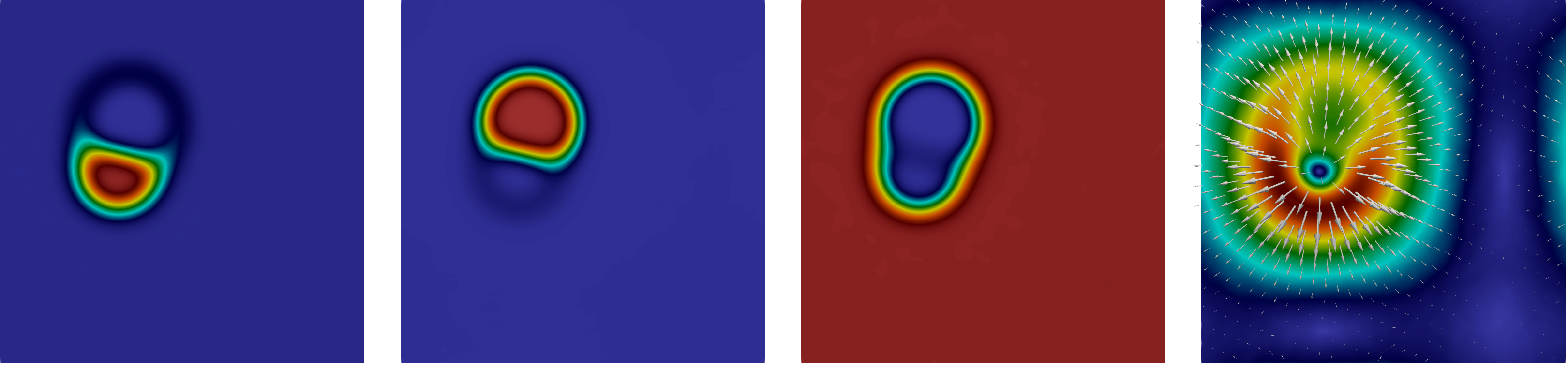} }}%
    \caption{Temporal evolution of the three phase fields together with the magnitud of the displacement vector and its direction.}
    \label{fig:evolution2}%
\end{figure}

From $t =5.64\times10^{-5}$, the system shows the merging of large inclusions and the action of the reversible chemical reaction (see Figure~\ref{fig:evolution2} (a)). At this stage, the creation of $\mathcal{A}^3$ predominates. One can verify such an assertion by checking the masses. However, there is no interfacial energy growth since the larger inclusions are merging. Therefore, the interfacial energies decrease. Moreover, the $\mathcal{A}^1$, $\mathcal{A}^2$, and $\mathcal{A}^3$ inclusions pass from rounded to square-like structures. Such behavior results from the dependency of the chemical potential upon the pressure $p^{\alpha}_{\varphi}$ caused by deformation and the Gibbs-Thomson effect associated with the curvature $\Delta \varphi^{\alpha}\R$. Along the boundaries of the inclusions, the driving force of the mass transport changes, which may generate inclusions of asymmetric morphologies. Moreover, the creation of $\mathcal{A}^1$, $\mathcal{A}^2$, and $\mathcal{A}^3$ following~\eqref{eq:chemical.reaction_ex.nchre.s} engenders a mechanical pressure associated with nucleation and growth. Figure~\ref{fig:evolution2} (a) also shows the large displacements at the phase boundaries that account for the impact of the chemical reaction on the stress generation (pressure). 

Figure~\ref{fig:evolution2} (b) shows the state at $t=4.26\times10^{-4}$. The system forms a chain-like structure composed of the phases $\mathcal{A}^1$ and $\mathcal{A}^2$, which is surrounded by the phase $\mathcal{A}^3$. This structure emerges as a result of the merging processes and the reversible chemical reaction. The reversible chemical reaction continues to take at the boundary between phases $\mathcal{A}^1$ and $\mathcal{A}^2$. Moreover, the phase $\mathcal{A}^3$ decomposes into $\mathcal{A}^1$ and $\mathcal{A}^2$. At this point in the evolution, the phase $\mathcal{A}^3$ composes almost the whole solid due to $k_{+} > > k_{-}$. The masses and interfacial energies for phases $\mathcal{A}^1$ and $\mathcal{A}^2$ decrease (see Figure~\ref{fig:masses.nchre}). However, for phase $\mathcal{A}^3$, the mass increases while reducing its interfacial energy (see Figure~\ref{fig:masses.nchre}). The phases are unmixed, whereby their concentrations correspond to the concentrations at the well points in the triple-well function. The displacements $u_x$ and $u_y$ are in the range of the previous stages. However, they move as the phases diffuse as a result of the relation between mass transport and deformation. Figure~\ref{fig:evolution2} (b) depicts the larger displacements are in line with the chain-like structure. 

Figure~\ref{fig:evolution2} (c) shows the evolution at $t=5.72\times10^{-4}$. The minimization of the global free energy as the system goes to the steady-state reduces the thickness of the chain-like structure. Eventually, the chain is composed of interleaved inclusions of phases $\mathcal{A}^1$ and $\mathcal{A}^2$. As the inclusions of $\mathcal{A}^1$ and $\mathcal{A}^2$ become smaller, the phase $\mathcal{A}^3$ encloses $\mathcal{A}^1$ and $\mathcal{A}^2$. The interfacial energies and masses keep decreasing since the action of the reversible chemical reaction has not ceased (see Figure~\ref{fig:masses.nchre}). The displacement field shows the interaction between the mechanical and chemical processes. We observe smaller displacements inside the chain-like structure and larger ones outside this region. 

In the time interval between $5.72\times10^{-4} < t < 1.84\times10^{-3}$, the inclusions of phases $\mathcal{A}^1$ and $\mathcal{A}^2$ shrink as the reversible reaction continues (see Figure~\ref{fig:evolution2} (d)). When the reversible chemical reaction ceases, the inclusions are rounded. The chemical reaction is active between $1\times10^{-5}<t<\times10^{-2}$. The larger (smaller) displacements appear around the inclusions of phase $\mathcal{A}^1$  ($\mathcal{A}^2$). These phases partially consume while the phase $\mathcal{A}^3$ gains mass. Finally, rounded structures composed of the three phases diffuse in the solid to generate displacements associated with mass transport. Figure~\ref{fig:evolution2} (e) portrays such a behaviour. Along with the whole evolution, free energy always behaves monotonically decreasing (see Figure~\ref{fig:freeener.nchre}).


\subsection{Ripening of Spherical Inclusions}

\begin{figure}
 \centering  
    \subfloat[]{{\includegraphics[scale=0.08]{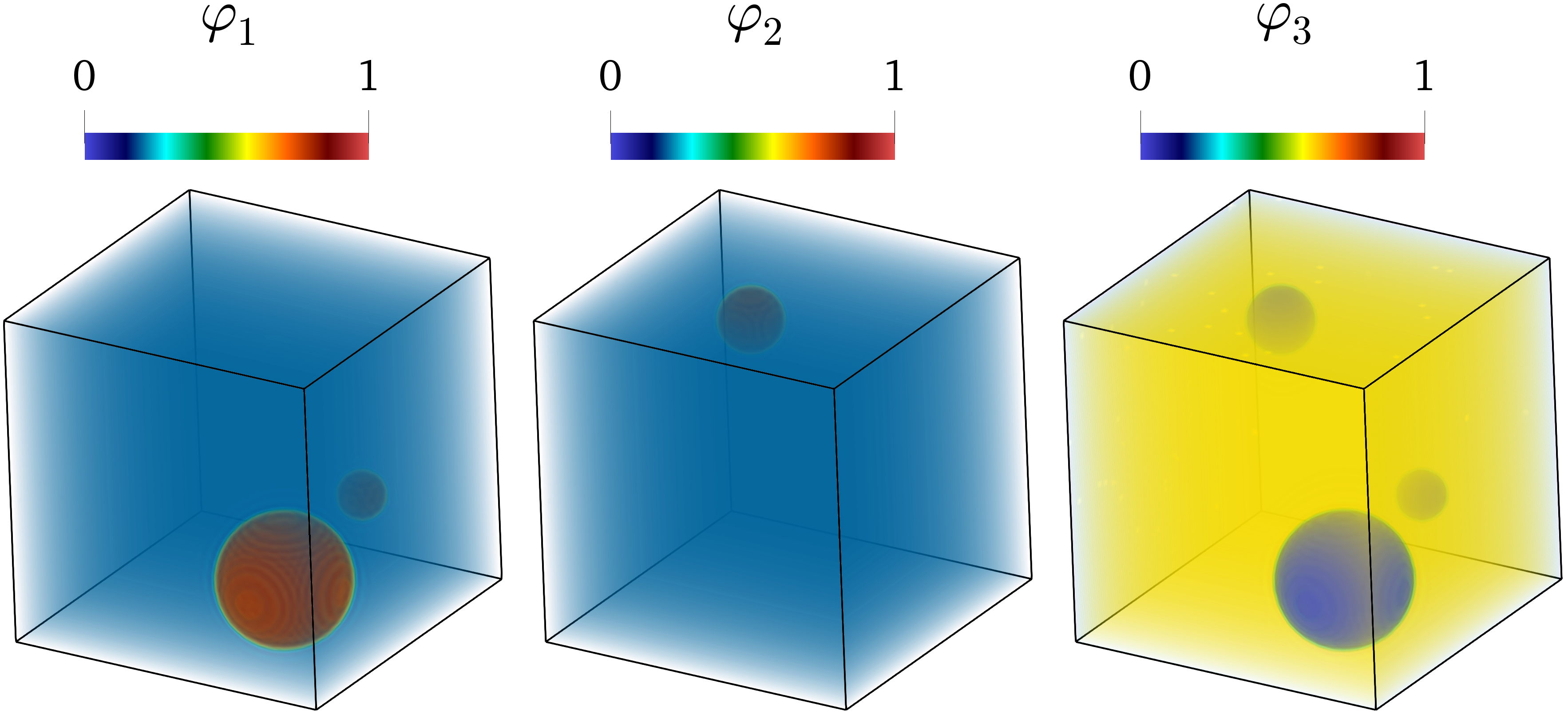} }}
     \includegraphics[scale=0.08]{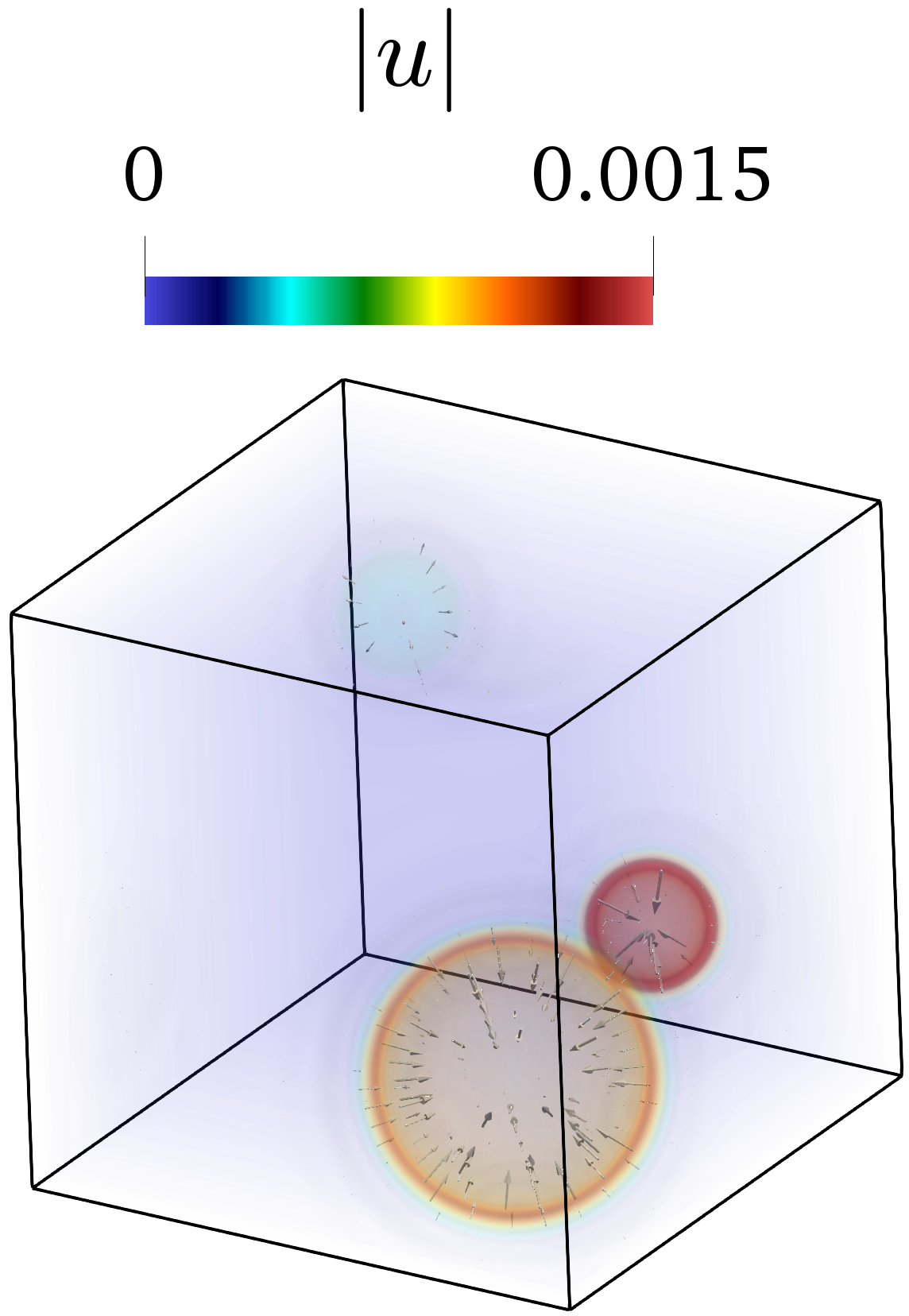}%
    \qquad
    \subfloat[]{{\includegraphics[scale=0.08]{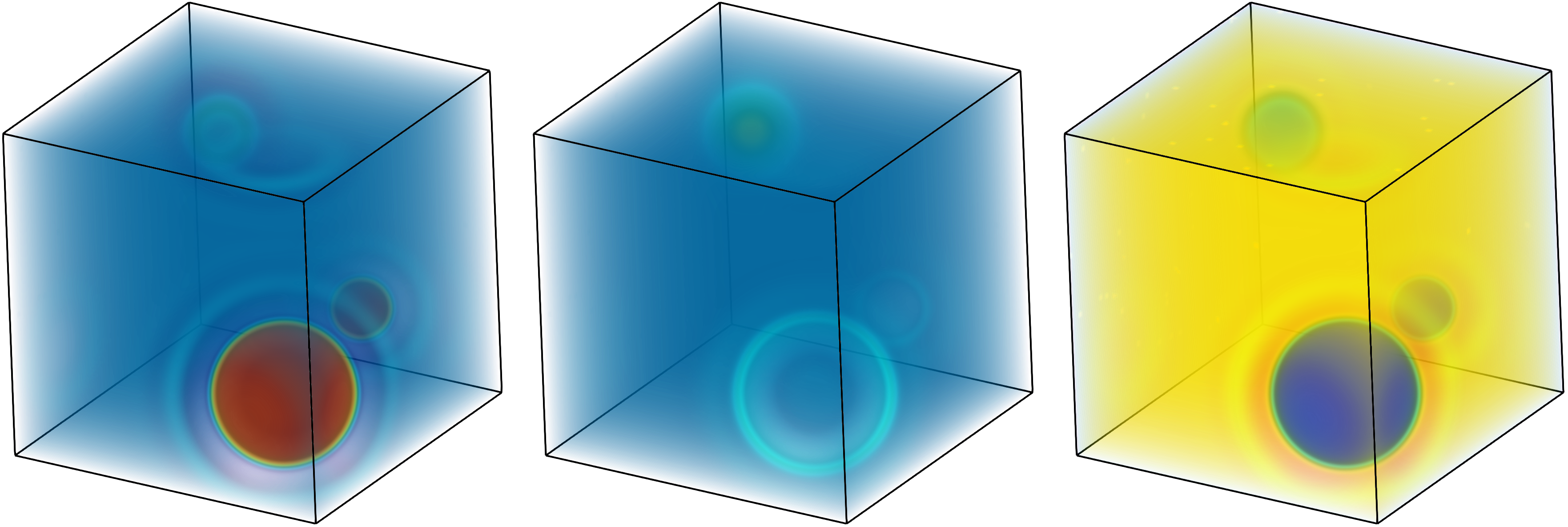} }}
    \includegraphics[scale=0.08]{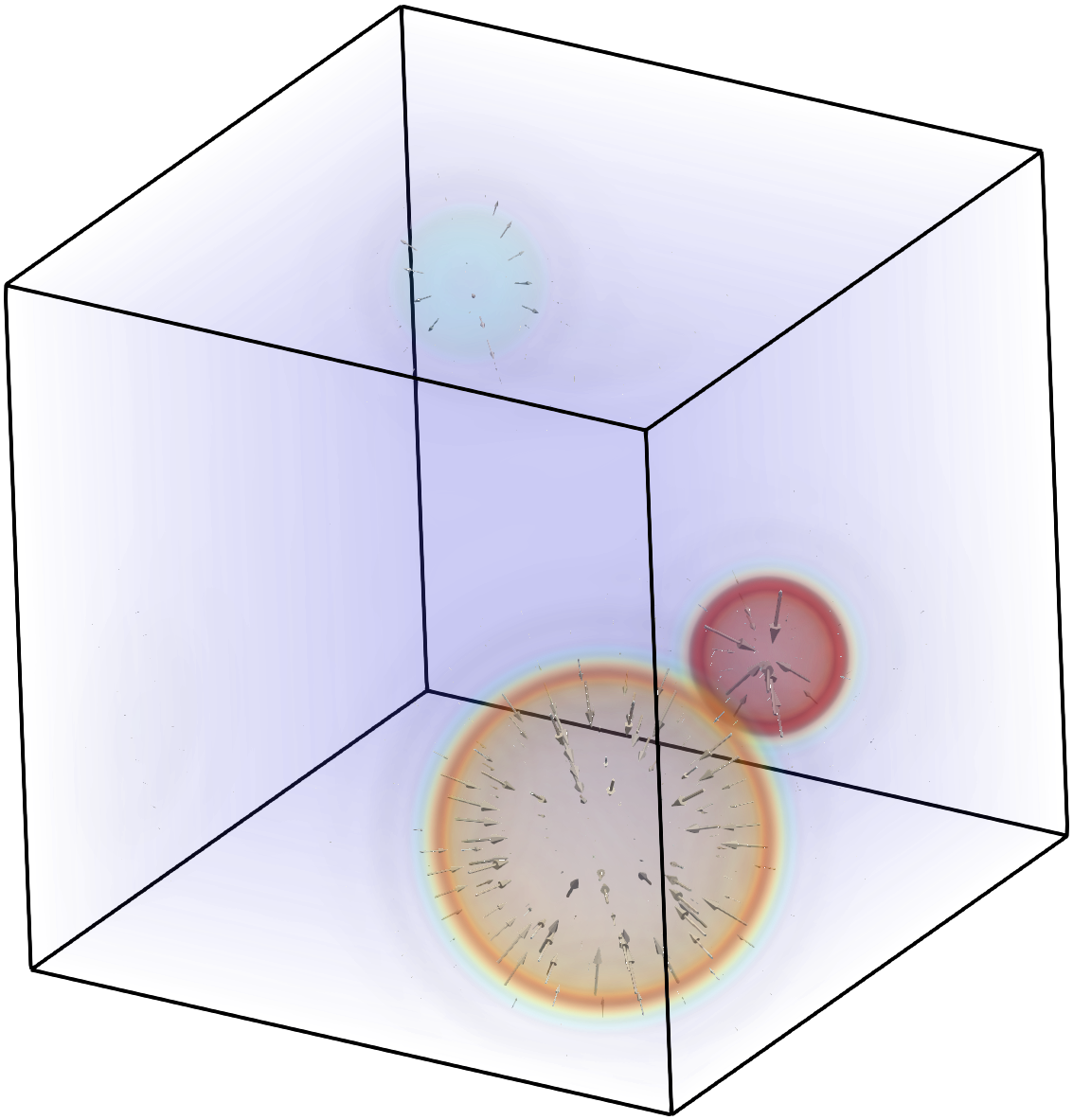}%
    \qquad
    \subfloat[]{{\includegraphics[scale=0.08]{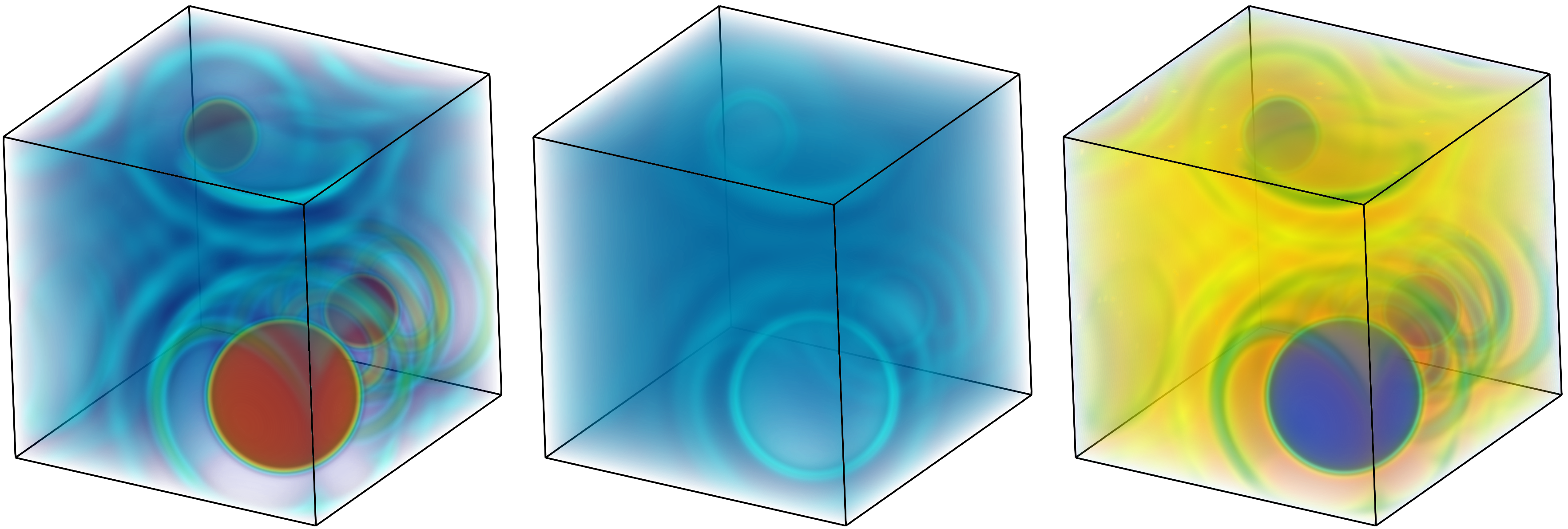} }}
    \includegraphics[scale=0.08]{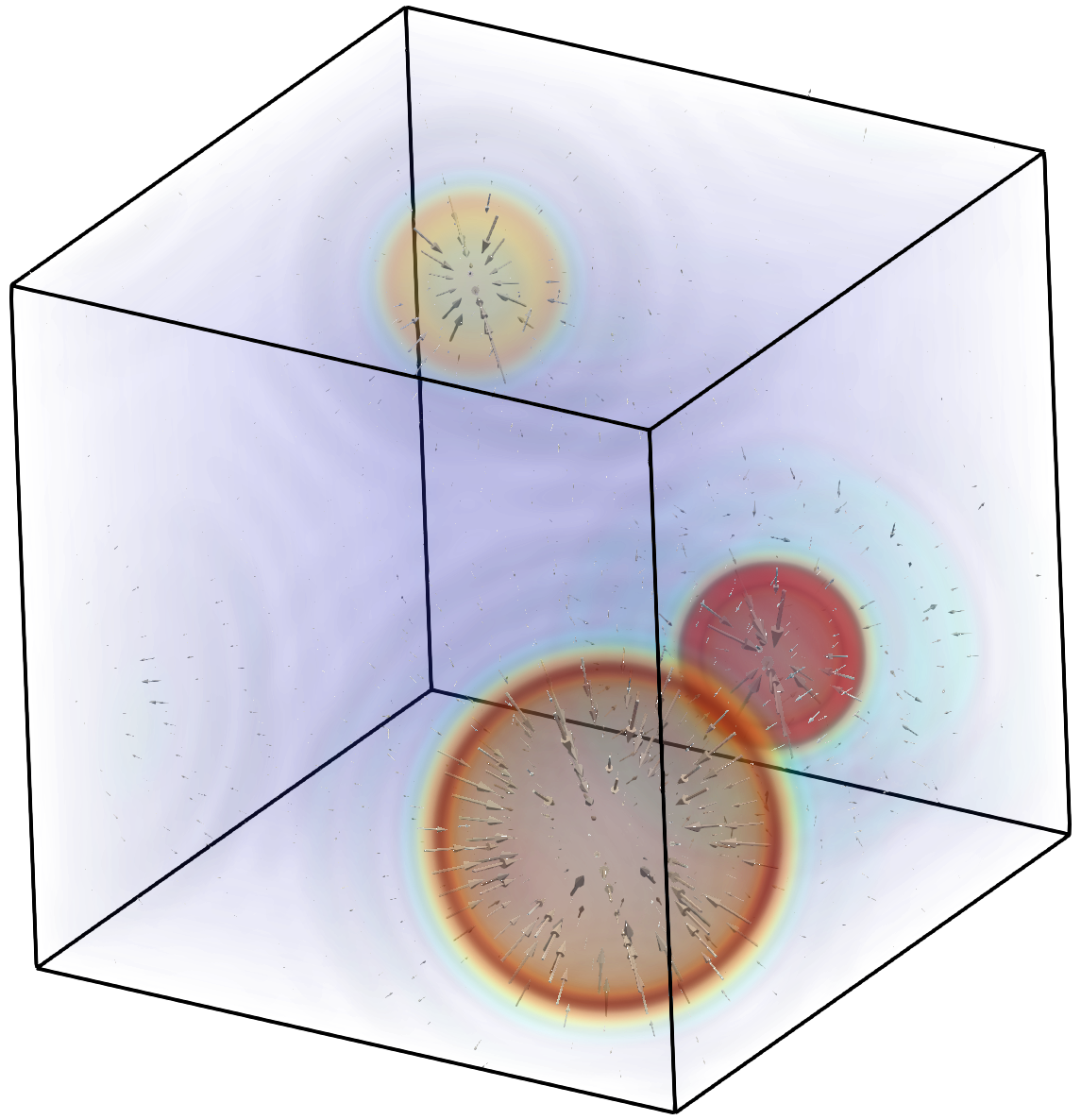}%
    \qquad
    \subfloat[]{{\includegraphics[scale=0.08]{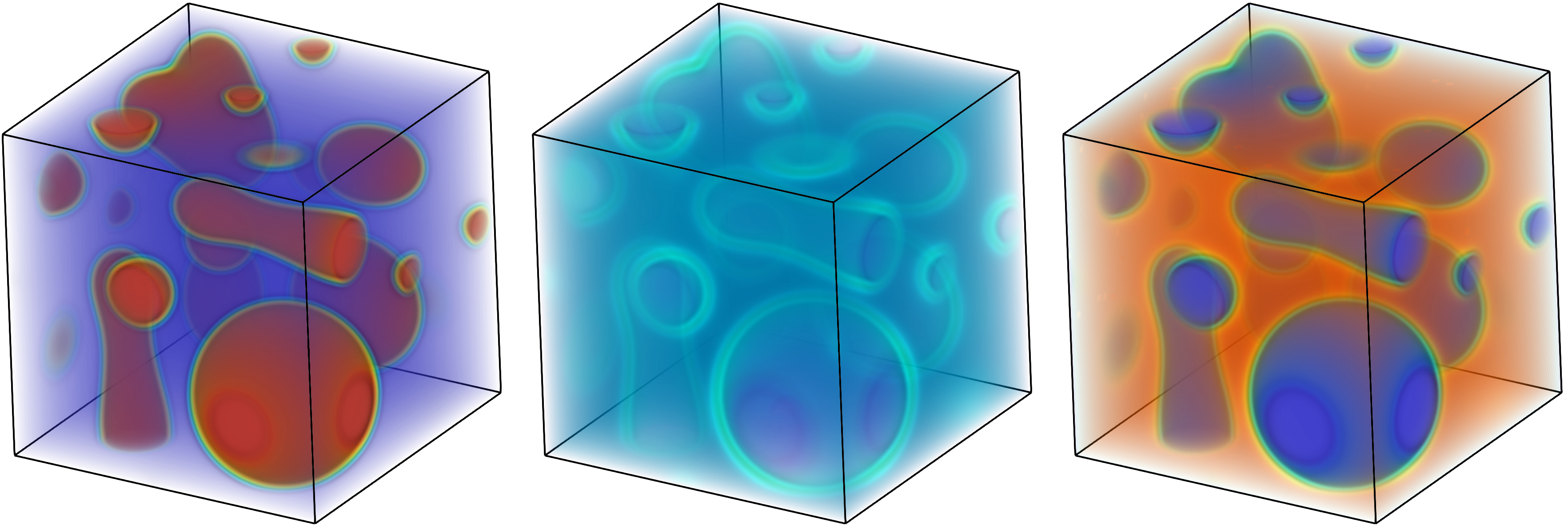} }}
    \includegraphics[scale=0.08]{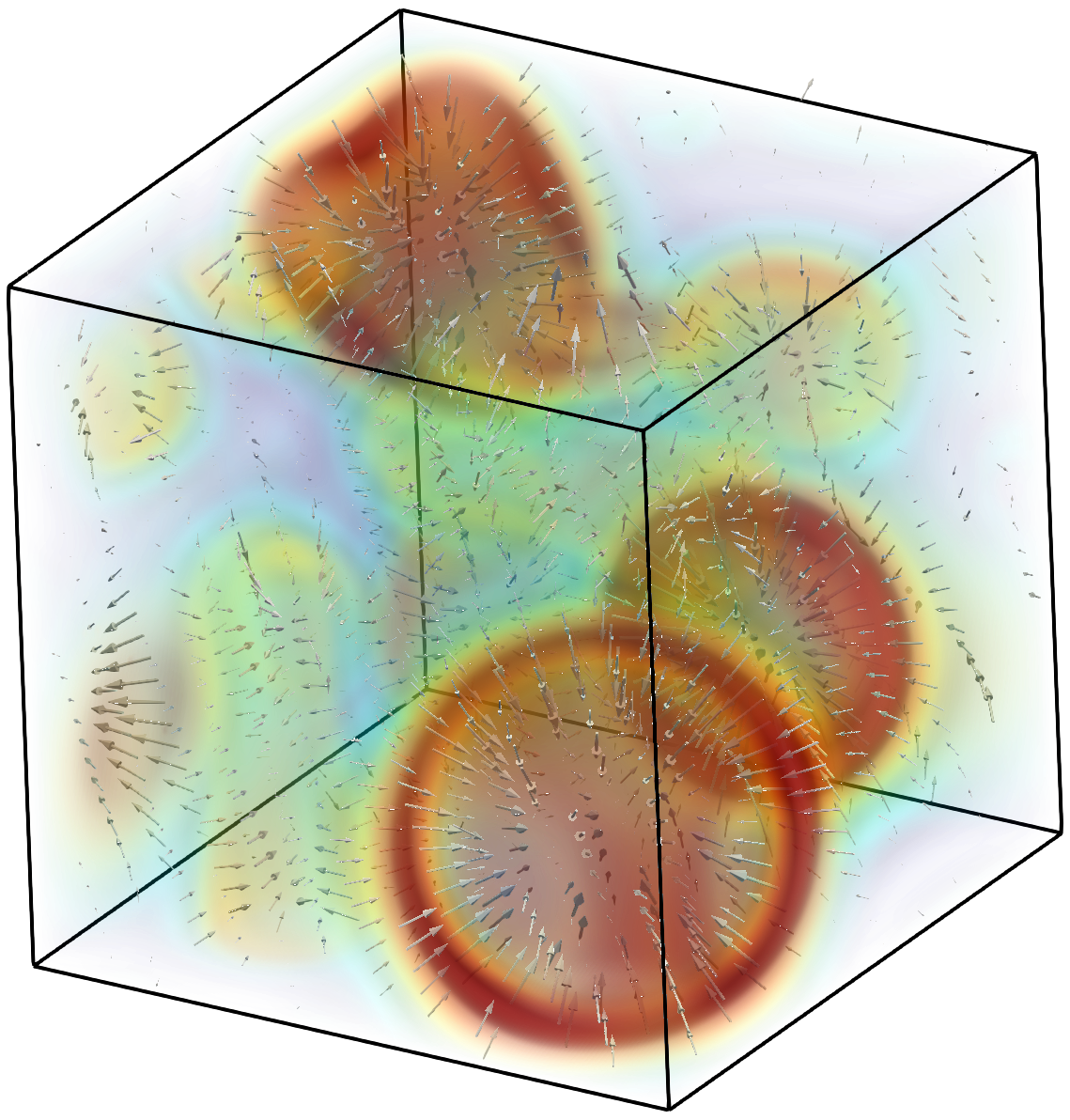}%
    \qquad
    \subfloat[]{{\includegraphics[scale=0.08]{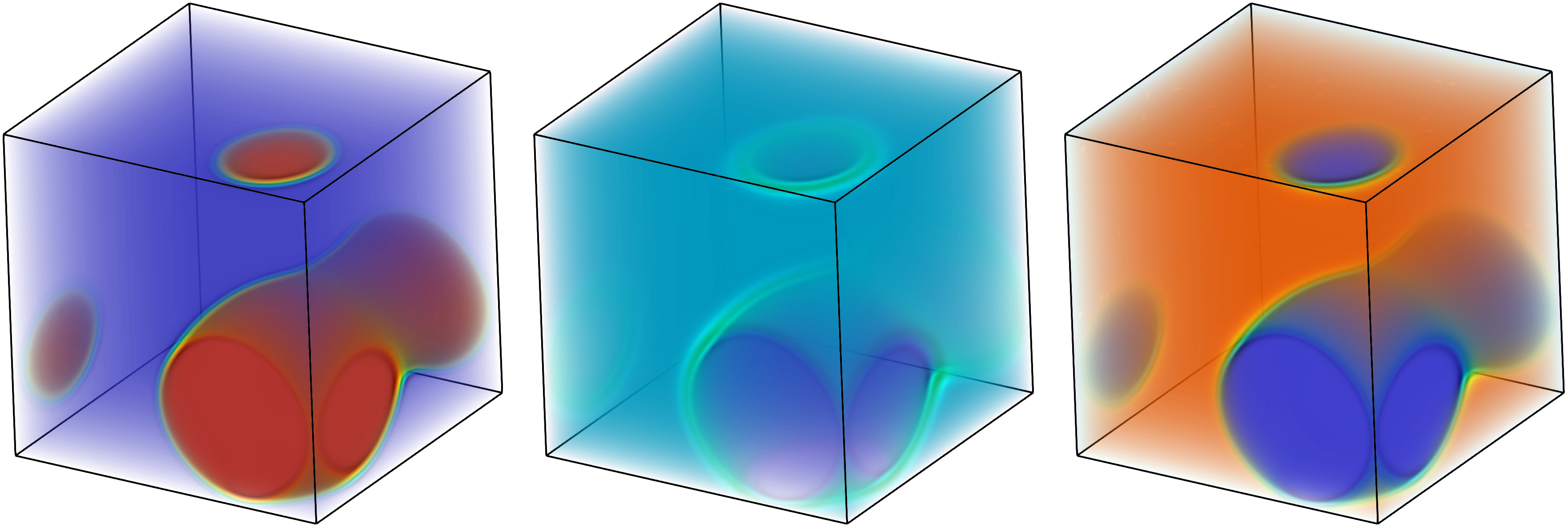} }}%
    \includegraphics[scale=0.08]{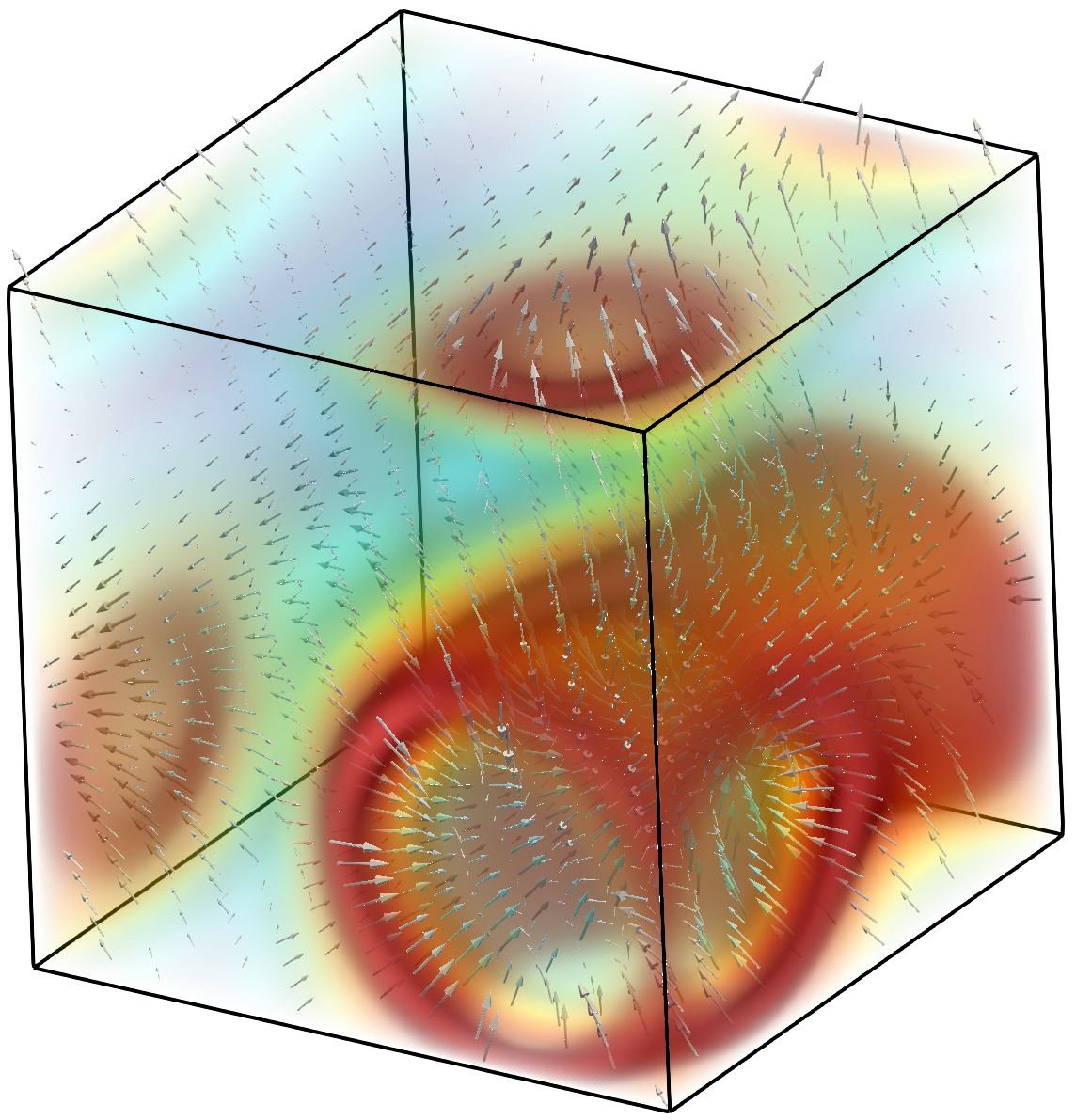}
    \caption{The inclusions differ in size which drives the ripening process. The deformation will result from the mass transport of the phases as the smaller inclusions go into the solution and deposit in the surface of the larger inclusions.}
    \label{fig:evo_3D.nchre}%
\end{figure}

We carry out a numerical simulation of a 3D configuration of three spherical inclusions. The spherical inclusions are composed of phases $\mathcal{A}^1$ and $\mathcal{A}^2$ while phase $\mathcal{A}^3$ serves as an interstitial phase. We study the stress-assisted volume changes triggered by the mass transport of the spherical inclusions associated with interfacial effects. We expect Ostwald ripening as a result of the differences in the size of the inclusions. We do not consider external contributions from body forces and external microforces. Consequently, we set $\textbf{b} = \textbf{0}$ and $\gamma^{\alpha} = 0$. Regarding the kinematics of the motion, we set $\bfv = \textbf{0}$ as we do not take into account inertial effects during these quasi-steady processes. We do not allow for chemical reactions between the phases, and therefore, the reactions rates $k_{+}$ and $k_{-}$ are zero. The latter entails that $s^{\alpha} = 0.0$. Hence, the stresses emerge solely from the mass transport associated with the interfacial interactions between the phases. Without loss of generality,  the initial condition serves as the reference configuration. We choose this reference state as an undeformed configuration of the body. The initial and boundary conditions are given by
\begin{equation}
\begin{aligned}
&S_{1} = (x-0.25)^2 + (y-0.25)^2 + (z-0.25)^2 - 0.2^2, \\
&S_{2} = (x-0.75)^2 + (y-0.75)^2 + (z-0.75)^2 - 0.1^2, \\
&S_{3} = (x-0.75)^2 + (y-0.75)^2 + (z-0.3)^2 - 0.08^2,\\
&h=0.2, \\
&\delta^1 = 0.31-0.8\left(0.5\tanh\left(\frac{S_{1}}{0.01h(h+2.0)} \right)+ 0.5\right),\\
&\delta^2 = 0.31-0.8\left(0.5\tanh\left(\frac{S_{2}}{0.01h(h+1.0)} \right)+ 0.5\right),\\
&\delta^3 = 0.31-0.8\left(0.5\tanh\left(\frac{S_{3}}{0.01h(h+0.8)} \right)+ 0.5\right),\\
&\varphi^1_0=1+\delta_1+\delta_2,\\
&\varphi^2_0=\delta_3,\\
&\varphi^3_0=1-\varphi^1_0-\varphi^2_0,\\
& \bfu = \boldsymbol{0},\\
&\text{in}\,\textbf{P},\text{subject to periodic boundary conditions}\text{on}\,\,\partial \textbf{P} \times(0,T).
\end{aligned}
\end{equation}
As in the previous examples, the range of the chemical and physical parameters are representative processes in geosciences. We use the same parameters as in Table~\ref{tb:pcparameters.nchre} in~\S\ref{sc:reversible.stresses}. Nevertheless, in this example we set the reaction rates $k_{+} = 0$ and $k_{-} = 0$ as mentioned above. The three phases diffuse with the same diffusion coefficient. Furthermore, the dimensionless numbers correspond to~\eqref{eq:dimensionless.parameters.nchre}. As mentioned before, the phase $\mathcal{A}^3$ serves as an interstitial phase following the mass constraint given by~\eqref{eq:const.nchre}. Figures~\ref{fig:evo_3D.nchre} (a) show the initial condition for the phases distribution and displacements, respectively. The system of equations to solve is given by
\begin{subequations}\label{eq:system.nchre.3d.N}
 \begin{align}
&\dot{\varphi} ^1  =  - \Divr\bfj^1_{\text{\tiny R}3}, \\
&\dot{\varphi} ^2  =  - \Divr\bfj^2_{\text{\tiny R}3}, \\
&\Divr \textbf{T}\R = \textbf{0}.
\end{align}    
\end{subequations}
where we use the phase  $\mathcal{A}^3$ as the reference species. We solve the system of partial differential equation~\eqref{eq:system.nchre.3d.N} in its primal form~\eqref{eq:weak.formulation.primal.material} and~\eqref{eq:weak.formulation.primal.material.l}. We state the problem as follows: find $\left\{ \varphi, \bfu\right\} \in \calC^2(\textbf{P})$ such that~\eqref{eq:nchre.system} given~\eqref{eq:gov.eq.nchre.N} subject to periodic boundary conditions up to the second derivative of $\varphi$, and $\bfu$ with respect to $\bfX$ in a cubic open region $\textbf{P}=(0,1)\times(0,1)\times(0,1)$. We use the PetIGA~\cite{ dalcin2016petiga} to solve the $64^3$ element mesh of a polynomial degree $2$ and continuity $1$. 

At early stages Figure~\ref{fig:evo_3D.nchre} (b), $t < 1.182\times10^{-6}$, part of the phase $\mathcal{A}^1$ deposits on the surface of the inclusion $\mathcal{A}^2$ as a new spherical inclusion $\mathcal{A}^1$ of same radius  appears. As the evolution proceeds, between the time range $1.182\times10^{-6} < t < 2.79\times10^{-6}$, the inclusion of phase $\mathcal{A}^2$ becomes smaller as its mass goes into the solution. There is no deposition of phase $\mathcal{A}^2$ at  this point in the evolution. Regarding the deformation, the stresses associated with the volume changes are small since the displacements do not change substantially (see Figure~\ref{fig:evo_3D.nchre} (c)). Nevertheless, after $t > 5.476\times10^{-6}$, rings composed of phases $\mathcal{A}^1$ and $\mathcal{A}^2$ appear around $\mathcal{A}^1$. This occurs as the solution gets supersaturated and the mass of phases $\mathcal{A}^1$ and $\mathcal{A}^2$ migrate to the surface of the more energetically stable structures in the system, which in this case correspond to the spherical  inclusions (see Figure~\ref{fig:evo_3D.nchre} (d)). Such mass transport induces volumetric stresses and concomitant displacements around the spherical inclusions (see Figure~\ref{fig:evo_3D.nchre} (d)). As the system tries to minimise its free energy, the masses of phases $\mathcal{A}^1$,  $\mathcal{A}^2$, and $\mathcal{A}^3$ in the solution separate and merge to form new spherical structures. The phase $\mathcal{A}^2$ locates around the spherical inclusions of $\mathcal{A}^1$. As mentioned before, the fact that the phases are diffusing induces volumetric stresses. Consequently, we see displacements where the phases are separating and merging (see Figure~\ref{fig:evo_3D.nchre} (d)). Later on, $t > 5.626\times10^{-3}$, the spherical inclusions composed of phase $\mathcal{A}^1$ and  $\mathcal{A}^2$ merge to form a more energetically stable distribution of elongated structures (see Figure~\ref{fig:evo_3D.nchre} (e)). Finally, the phase $\mathcal{A}^2$ wraps the phase $\mathcal{A}^1$, and $\mathcal{A}^3$ acts as an interstitial phase. The structure at steady state emerges as a result of the coupled chemo-mechanical interactions of the three-component system where the source of stress generation solely results from the mass transport of the phases (see Figure~\ref{fig:evo_3D.nchre} (e)). 

\section{Conclusions}

We develop a thermodynamically-consistent model that describes the evolution of chemically active mineral solid solutions. The theoretical foundations of the framework rely on modern continuum mechanics, thermodynamics far from equilibrium, and the phase-field model, which allow us to derive a set of coupled chemo-mechanical partial differential equations. Using 2D and 3D numerical simulations, we show the evolution of the stress-assisted volume changes triggered by mass transport and chemical reactions. We also include a constraint system using the Larch\'e--Cahn derivative, which takes into account the mass constraint imposed by the solid crystalline structure. Finally, we conclude that the influence of mechanical effects upon chemically active mineral solid solutions must be taken into account and most importantly, they play a substantial role in the evolution of these geosystems. 

\section{Acknowledgments}

We are indebted to Professor Eliot Fried. We had many exhaustive discussions in which he gave us valuable ideas, constructive comments, and encouragement. This publication was made possible in part by the CSIRO Professorial Chair in Computational Geoscience at Curtin University and the Deep Earth Imaging Enterprise Future Science Platforms of the Commonwealth Scientific Industrial Research Organisation, CSIRO, of Australia. The European Union's Horizon 2020 Research and Innovation Program of the Marie Sk\l{}odowska-Curie grant agreement No. 777778, and the Mega-grant of the Russian Federation Government (N 14.Y26.31.0013) provided additional support. Lastly, we acknowledge the support provided at Curtin University by The Institute for Geoscience Research (TIGeR) and by the Curtin Institute for Computation.

\appendix

\section{Thermodynamics and Continuum Mass and Forces}\label{ap:framework}

\subsection{Measure of Strain}

In deforming bodies undergoing mass transport and chemical reactions, the particles move relative to each other as a result of external forces and compositional changes. A description of this movement measures the relative displacement of the particles. We use a Lagrangian description of the displacement field $\bfu$ which defines the kinematics of the motion, that is, 
\begin{equation}
\bfu = \bfx (\bfX, t)- \bfX,
\end{equation}
and the deformation gradient 
\begin{equation}
\textbf{F} = \nabla \bfchi_{t} = \nabla \bfu + \textbf{I},
\end{equation}
where \textbf{I} denotes the second-order identity tensor. To ensure an admissible deformation, that is, a continuum body cannot penetrate itself, the Jacobian of the deformation gradient must fulfill the following constraint
\begin{equation}
J \Def \text{det} \ \textbf{F} > 0.
\end{equation}
The velocity of a material particle $\bfX$ as a function of the motion is
\begin{equation}
\bfV \Def \frac{\partial \bfchi (\bfX, t)}{\partial t},
\end{equation}
and its counterpart in the current configuration is
\begin{equation}\label{eq:vel}
\bfv \Def \frac{\partial \bfchi (\bfX, t)}{\partial t}  \Big\vert_{\bfX=\bfchi^{-1} (\bfx, t)}.
\end{equation}
In~\eqref{eq:vel}, we describe the velocity at time $t$ of a material particle located at $\bfx = \bfchi_{t} (\bfX)$. \\

Given the definition of the deformation gradient and the spatial velocity, the right Cauchy-Green stress, the Green-Lagrange strain, the rate of strain tensors and the spatial velocity gradient are given by 
\begin{equation}
\textbf{C} = \textbf{F}^{\trans} \textbf{F},
\end{equation}
\begin{equation}
\textbf{E} = \frac{1}{2} \bigg(\textbf{F}^{\trans} \textbf{F} - \textbf{I}\bigg) = \frac{1}{2} (\textbf{C} - \textbf{I}),
\end{equation} 
\begin{equation}
\textbf{D} = \sym (\grad \bfv) = \frac{1}{2} ( \grad \bfv +  \grad \bfv^{\trans} ),
\end{equation} 
\begin{equation}\label{eq:vel.grad}
\textbf{L} =  (\grad \bfv) = \dot{\textbf{F}}\textbf{F}^{-1}.
\end{equation} 
We apply the change of variables theorem to relate the reference and current configurations. In particular, these equations account for the deformation of an infinitesimal line, area, and volume element, that is,
\begin{equation}
\dx = \textbf{F} \dxr,
\end{equation} 
\begin{equation}\label{eq:dadA}
\da = J \textbf{F}^{-\trans} \dar,
\end{equation} 
\begin{equation}\label{eq:transvol}
\dv = J \dvr.
\end{equation} 

\subsection{Fundamental Balances}

We derive a set of balance equations in the form of partial differential equations that define how the mass, linear and angular momenta, internal energy, and entropy vary in time as the solid-species system endures mechanical and chemical processes. As suggested in~\cite{ gurtin2010mechanics, dal2015computational, miehe2016phase, tsagrakis2017thermodynamic}, three primary fields govern the coupled chemo-mechanical responses of the solid: the deformation $\bfchi(\textbf{X}, t)$, the species concentration $\varphi^{\alpha}\R(\textbf{X}, t)$ per unit of reference volume, and the chemical potential $\mu^{\alpha}\R(\textbf{X}, t)$ per unit of reference volume where $\alpha$ denotes the $\alpha$-th species that composes the solid.

Let $\textbf{P} \subset \textbf{B}$ be an arbitrary control volume in conjunction with its boundary $\text{S} = \partial \textbf{P}$. Analogously, consider $\calP_{t}$ as a bounded control volume of $\calB_{t}$ such that $\calP_{t} = \bfchi (\textbf{P})$ with boundary $\calS = \partial \calP_{t}$. According to Cauchy's theorem, the traction $\textbf{t}$ on a surface $\da \subset \calS$ and whose normal $\bfn$ points outwards is $\bft = \textbf{T} (\bfx, t) \bfn $. This traction characterizes the force exerted by the rest of the body $\calB_{t} \setminus \calP_{t}$ on $\calP_{t}$ through  $ \da \subset \calS$~\cite{ dal2015computational, miehe2016phase}. The traction \textbf{t} depends linearly pointwise on the normal $\bfn$ through Cauchy's stress tensor \textbf{T}~\cite{ gonzalez2008first}. Applying the Equation~\eqref{eq:dadA} to the identity $\textbf{t}_{\text{\tiny R}}  \dar = \textbf{t} \da$, we find the force acting on the surface element $\da$ as a function of the surface element $\dar$~\cite{ dal2015computational, gurtin2010mechanics}. This identity leads to the nominal stress tensor $\textbf{T}_{\text{\tiny R}}$, that is, the first Piola-Kirchhoff, 
\be
\textbf{T}\R \dar = \textbf{T} \da \ \ \ \ \text{with} \ \ \ \ \textbf{T}_{\text{\tiny R}} = J \textbf{T} \textbf{F}^{-\trans}.
\ee

As mentioned above, the chemo-mechanical interactions take place through an elastically deforming solid composed by a network and constituent species. Consequently, we formulate balances of mass conservation for both the solid and the constituent species. Thus, we define $\varphi^{\alpha}\R$ as the local concentration of the $\alpha$-th species per unit of undeformed configuration together with a spatial species outflux $\bfj^{\alpha}$. In agreement with the balance of mass conservation, the rate of mass change of the $\alpha$-th species in the control volume $\textbf{P}$ has to be equal to the contribution from the mass supply, typically caused by chemical reactions between the species, and the net mass flux through the boundary $\calS$, that is, 
\be \label{eq:massbal1}
\dot{\overline{\int_{\textbf{P}} \varphi^{\alpha}\R \dvr}} = \int_{\textbf{P}} s^{\alpha} \ \dvr - \int_{\calS} \bfj^{\alpha} \cdot \bfn \, \text{d}v,
\ee
where $s^{\alpha} $ is the mass supply expressed in the reference configuration. The mass supply is composed of two terms, an external contribution due to external agents and internal contributions caused by chemical reactions. Thereby,
\begin{equation}\label{eq:mass.supply.nchre}
s^\alpha = s^\alpha_{\text{int}} + s^\alpha_{\text{ext}}.
\end{equation}

Using the divergence theorem, we transform the surface integral of the species flux into a volume integral of the divergence of the species flux as follows 
\be\label{eq:massbalR2.nchre} 
\dot{\overline{\int_{\textbf{P}} \varphi^{\alpha}\R \dvr}} = \int_{\textbf{P}} s^{\alpha} \ \dvr - \int_{\calS} \Divc \bfj^{\alpha}  \, \text{d}v.
\ee
The Lagrangian description of Equation~\eqref{eq:massbalR2.nchre} is
\be\label{eq:massbal3}
\dot{\overline{\int_{\textbf{P}} \varphi^{\alpha}\R \dvr}} = \int_{\textbf{P}} s^{\alpha} \ \dvr - \int_{\textbf{P}} \Divr \bfj^{\alpha}\R  \, \text{d}v_{\text{\tiny R}},
\ee
where we use the Piola transform. Thus, the material species flux is then $\bfj^{\alpha}\R = \textbf{F}^{-1} (J \bfj^{\alpha})$. Finally, the localized version of Equation~\eqref{eq:massbal3} is given by
\be
\dot{\varphi^{\alpha}\R} = s^{\alpha}  - \Divr \bfj^{\alpha}\R.
\ee
The concentration of each species is linearly dependent on the other, via the following constraint,
\be\label{eq:const.nchre}
\sum _{\alpha=1} ^{n} \varphi^{\alpha}\R = 1,
\ee
which renders
\be\label{eq:gradconst.nchre}
\sum _{\alpha=1} ^{n} \dot{\varphi^{\alpha}\R} = 0 \ \ \text{and} \ \ \sum _{\alpha=1} ^{n} \nabla \varphi^{\alpha}\R = 0,
\ee
where $n$ stands for the total number of species. The mass constraint that Equation~\eqref{eq:const.nchre} expresses must hold when the solid is solely composed of the diffusing species. Herein, we restrict our attention to the case where mass transport by vacancies is not feasible.

Henceforth, a superimposed dot $(\ \dot{} \ )$ stands for the material time derivative, for instance, $\dot{\varphi^{\alpha}\R}$ is the material time derivative of the concentration species. Given the conservation of the solid mass, we define $\rho$ and $\rho_{0}$ as the solid density in the current and reference configuration, respectively. Then, the balance of solid mass reads
\be\label{eq:smass}
\int_{\calP_{t}} \rho \, \text{d}v = \int_{\textbf{P}} \rho_{0}  \, \text{d}v_{\text{\tiny R}},
\ee
In ~\eqref{eq:smass}, we convert the volume integral in the current configuration into its counterpart in the reference configuration by employing the relation~\eqref{eq:transvol}. Finally, we use the localization theorem that leads to the local conservation of solid mass 
\be\label{eq:smass1}
\rho_{0}=  J\rho.
\ee
Neglecting all inertial effects; namely, we assume the spatial velocity $\bfv$ is invariant through the time, the balance of conservation of linear momentum reads
\be\label{eq:blm}
\int_{\calS} \bft \da + \int_{\textbf{P}} \textbf{b}  \, \text{d}v_{\text{\tiny R}} = \mathbf{0}. 
\ee
The balance of linear momentum relates forces to changes in the motion of the body. Such balance involves the traction $\textbf{t}$ acting on a surface element $\da$ as well as a body force $\textbf{b}$. Conventionally, the body force  $\textbf{b}$ accounts for forces resulting from gravitational effects. Through the divergence theorem and the definition~\eqref{eq:trac.cauchy}, we express the surface integral in ~\eqref{eq:blm} as a volume integral 
\be\label{eq:blm1}
\int_{\calP_{t}} \Divc \textbf{T} \dv + \int_{\textbf{P}} \textbf{b} \dvr = \mathbf{0},
\ee
and after some straightforward manipulations in ~\eqref{eq:blm1}, the localized Lagrangian form of the balance of linear momentum is 
\be\label{eq:lm}
\Divr \textbf{T}\R + \textbf{b} = \mathbf{0}.
\ee
The balance of conservation of angular momentum is 
\be
\int_{\calP_{t}} \bfx \times \bft \dv + \int_{\textbf{P}} \bfx \times \textbf{b} \dvr = \textbf{0}.
\ee
After using the definition of the balance of linear momentum, the divergence theorem, and the localization theorem, we end up with $\textbf{T}^{\trans} = \textbf{T}$ . The previous relation implies the symmetry of the Cauchy's tensor~\cite{ gurtin1982introduction, sedov1972course}. Finally, the localized Lagrangian form of the balance of angular momenta is
\be
\text{skw} \textbf{T}\R \textbf{F}^{\trans} = \textbf{0}.
\ee
Following the line of thought introduced by Gurtin and Fried~\cite{ gurtin1996generalized, miranville2001consistent, bonfoh2001cahn, cherfils2011cahn}, we separate balances of conservation laws from constitutive equations. As a consequence, we include a balance of microforces, that is 
\be
\int_{\textbf{P}} (\pi^\alpha + \gamma^\alpha) \dvr =  - \int_{\calS} \bfxi^{\alpha} \cdot \bfn \da,
\ee
where the vector $\bfxi^{\alpha}$ and  the scalar $\pi^\alpha$ ($\gamma^\alpha$) correspond to the $\alpha$-th microstress and $\alpha$-th the internal (external) microforce, respectively. In general, the microstresses and microforces are quantities associated with microscopic configurations of atoms. We express the balance of microforces in a Lagrangian form
\be
\int_{\textbf{P}} (\pi^\alpha + \gamma^\alpha) \dvr = - \int_{\textbf{P}} \Divr \boldsymbol{\xi}^{\alpha}_{\text{\tiny R}} \dvr,
\ee
and after applying the locatization theorem, the balance for microforces reads
\be\label{eq:balmicro1}
\pi^\alpha + \gamma^\alpha = - \Divr \boldsymbol{\xi}^{\alpha}_{\text{\tiny R}},
\ee
where $\boldsymbol{\xi}^{\alpha}_{\text{\tiny R}} = \textbf{F}^{-1} (J \bfxi^{\alpha})$.

\subsection{Laws of Thermodynamics and Free Energy Inequality}

We separate conservation statements from constitutive equations as suggested by Gurtin \& Fried~\cite{ gurtin1996generalized, miranville2001consistent, bonfoh2001cahn, cherfils2011cahn}. To describe the thermodynamics of this system, we introduce a power expenditure $\calW_{\text{ext}} = \calW_{\text{ext}}(\textbf{P}) + \calW_{\text{ext}}(\prt)$ externally to $\textbf{P}$ and $\prt$ done by the external microforce and force on $\textbf{P}$, and the microtraction and traction on $\calS$
\begin{subequations}\label{eq:external.power.nchre}
\begin{align}
\calW_{\text{ext}}(\textbf{P})&=\sum_{\alpha=1}^{n}\left\{\int\limits_{\textbf{P}}\gamma^\alpha\dot{\varphi}^\alpha\R\dvr\right\} + \int\limits_{\textbf{P}}\textbf{b} \cdot \bfv \dvr, \\
\calW_{\text{ext}}(\prt)&=\sum_{\alpha=1}^{n}\left\{\int\limits_{\calS}\xis^\alpha\dot{\varphi}^\alpha\R \da\right\} + \int\limits_{\calS} \bft \cdot \bfv \da.
\end{align}
\end{subequations}
By neglecting all inertial effects and body forces,  we use the first law of thermodynamics to characterize the balance between the rate of internal energy $\dot{\varepsilon}$ and the expenditure rate of the chemo-mechanical power, caused by external forces, species transport, and chemical reactions. The first law is then,  
\be\label{eq:firslawt.nchre}
\dot{\overline{\int_{\textbf{P}} \varepsilon \dvr}} =  \calW_{\text{ext}} -\int_{\calS} \bfq \cdot \bfn \ \da + \int_{\textbf{P}} r \dvr - \sum_{\alpha=1}^{n}  \left\{\int_{\calS} \mu^{\alpha}\R\bfj^{\alpha} \cdot \bfn \da - \int_{\textbf{P}} \mu^{\alpha}\R s^{\alpha}_{\text{ext}} \dvr \right\}.
\ee
There is no contribution of $s^{\alpha}_{\text{int}}$ to the energy balance~\eqref{eq:firslawt.nchre}. The entropy imbalance, in the form of the Clausius--Duhem inequality, states that the rate of growth of the entropy $\eta$ is at least as large as the entropy flux $\bfq / \vartheta$ plus the contribution from the entropy supply $q/\vartheta$, that is,
\be\label{eq:seclawt}
\dot{ \overline{\int_{\textbf{P}} \eta \dvr}} \geq - \int_{\calS} \frac{\bfq \cdot \bfn}{\vartheta} \da + \int_{\textbf{P}} \frac{r}{\vartheta} \dvr,
\ee
where $\bfq$, $r$, and $\vartheta$ stand for the spatial heat flux, heat supply, and temperature, respectively. 
The localized Lagrangian version of~\eqref{eq:firslawt.nchre} and~\eqref{eq:seclawt} read
\be\label{eq:firslawt2}
\dot{\varepsilon} =  \calW_{\text{ext}} - \Divr \bfq\R + r - \sum_{\alpha=1}^{n} \left\{\Divr \mu^{\alpha}\R \bfj^{\alpha}\R - \mu^{\alpha}\R s^{\alpha}_{\text{ext}}\right\},
\ee
 and 
\be\label{eq:seclawt2}
\dot{\eta} \geq - \Divr \vartheta^{-1} \bfq \R + \vartheta^{-1} r,
\ee
where $\bfq \R = \textbf{F}^{-1} (J \bfq)$ is the material heat flux. Moreover, $\calW_{\text{ext}}$ is
\begin{equation}
\calW_{\text{ext}} =\sum_{\alpha=1}^{n} \left\{ ( \Divr \boldsymbol{\xi}^{\alpha}_{\text{\tiny R}}  + \gamma^{\alpha} ) \dot{\varphi}^\alpha\R + \boldsymbol{\xi}^{\alpha}_{\text{\tiny R}} \cdot \nabla \dot{\varphi}^\alpha\R\right\} + (\Divr \textbf{T}\R + \textbf{b})\cdot\bfv + \textbf{T}\R \colon \dot{\textbf{F}} \label{eq:external.power.nchre.17}.
\end{equation}
 Rewriting~\eqref{eq:firslawt2} and~\eqref{eq:seclawt2}, and multiplying~\eqref{eq:seclawt2} by $\vartheta$, we obtain
\be
\dot{\varepsilon} = \calW_{\text{ext}} - \Divr \bfq\R + r - \sum_{\alpha=1}^{n} \left\{\nabla \mu^{\alpha}\R \cdot \bfj^{\alpha}\R + \mu^{\alpha}\R \Divr \bfj^{\alpha}\R - \mu^{\alpha}\R s^{\alpha}_{\text{ext}} \right\},
\ee
 and
\be
\vartheta \dot{\eta} \geq \vartheta^{-1} \nabla \vartheta \cdot \bfq\R - \Divr \bfq\R  + r.
\ee
The Helmholtz free energy results from applying the Legendre transform to the internal energy while replacing the entropy of the system by the temperature as an independent variable., i.e., $\dot{\psi} = \dot{\varepsilon} - \dot{\vartheta} \eta - \vartheta \dot{\eta}$. Consequently, we obtain 
\be\label{eq:localfreener}
\dot{\psi} \leq  \calW_{\text{ext}} - \sum_{\alpha=1}^{n} \left\{\nabla \mu^{\alpha}\R \cdot \bfj^{\alpha}\R + \mu^{\alpha}\R \Divr \bfj^{\alpha}\R - \mu^{\alpha}\R s^{\alpha}_{\text{ext}}\right\} - \vartheta^{-1} \nabla \vartheta \cdot \bfq\R - \dot{\vartheta}\eta,
\ee
Introducing the balances of both mass conservation and microforces into~\eqref{eq:localfreener}, the free energy imbalance under isothermal conditions is 
\be\label{eq:energyimb}
\dot{\psi} \leq  \textbf{T}\R \colon \dot{\textbf{F}} + \sum_{\alpha=1}^{n} \left\{(\mu^{\alpha}\R - \pi^{\alpha}) \dot{\varphi^{\alpha}\R} + \boldsymbol{\xi}^{\alpha}_{\text{\tiny R}} \cdot \nabla \dot{\varphi^{\alpha}\R} - \bfj^{\alpha}\R \cdot \nabla \mu^{\alpha}\R - \mu^{\alpha}\R s^{\alpha}_{\text{int}}\right\}.
\ee

\subsection{The Principle of Material Frame Indifference}

Throughout the derivation of the constitutive behavior of the multicomponent solid, we use the Larch\'e--Cahn derivative for both scalar and gradient fields as expressed by \cite{Cla21a}, together with the mass constraint given by~\eqref{eq:const.nchre}. We assume the following constitutive dependence of the free energy $\psi$
\be\label{eq:constfree.nchre}
\psi = \hat{\psi}(\bfvarphi\R, \nabla \bfvarphi\R, \textbf{F}) = \hat{\psi}^{ch}(\bfvarphi\R, \nabla \bfvarphi\R) + \hat{\psi}^{el}(\textbf{\textbf{F}}^{e}(\textbf{F},\bfvarphi\R)).
\ee
The objectivity principle requires the constitutive relation~\eqref{eq:constfree.nchre} to be invariant under a superposed rigid body motion or equivalently, independent of the observer. We can relate two different displacement fields $\chi$ and $\chi^{*}$ as follows 
\be\label{eq:rigidbody}
\chi^{*} (\textbf{X}, t) = \textbf{Q}(t)\chi(\textbf{X},t) + \textbf{c}(t),
\ee
where $\textbf{Q}(t)$ represents a rotation tensor and $\textbf{c}(t)$ the relative translations. Therefore, the transformation of the potential~\eqref{eq:constfree.nchre} following~\eqref{eq:rigidbody} implies
\be\label{eq:constfree1.nchre}
\psi = \hat{\psi}(\bfvarphi\R, \nabla \bfvarphi\R, \textbf{F}) = \overline{\psi}(\bfvarphi\R, \nabla \bfvarphi\R, \textbf{C}),
\ee
which ensures consistency with the dissipation inequality~\eqref{eq:energyimb} and the principle of frame-indifference.

\bibliographystyle{elsarticle-num}

\end{document}